\documentclass[floatfix, noshowpacs, preprintnumbers, twocolumn, amsmath, amssymb, aps, pra, superscriptaddress]{revtex4-1}

\usepackage{times}
\usepackage{array}
\usepackage{amsmath}
\usepackage{amsfonts}
\usepackage{amssymb}
\usepackage{amsthm}
\usepackage{amsbsy}
\usepackage{stmaryrd}
\usepackage{pifont}
\usepackage{latexsym}
\usepackage{mathtools}
\usepackage{bbold}
\usepackage{gensymb}
\usepackage[all]{xy}
\usepackage{graphicx}
\usepackage{braket}
\usepackage[T1]{fontenc}
\usepackage{booktabs}
\usepackage{enumerate}
\usepackage{soul}

\usepackage{epstopdf}
\usepackage{epsfig}
\usepackage[labelformat=default, labelsep=default, justification=centerlast,  font=footnotesize]{caption}
\usepackage{verbatim}
\usepackage{multirow}
\usepackage{rotating}
\usepackage{minibox}
\usepackage{tabularx}

\usepackage[titletoc,toc,title]{appendix}
\usepackage{enumerate}
\usepackage{enumitem}
\usepackage{color}
\usepackage[dvipsnames]{xcolor}

\usepackage{comment}
\usepackage{sidecap}
\usepackage{bbm}
\usepackage{bm}
\usepackage{braket}
\usepackage{float}

\usepackage[normalem]{ulem}

\DeclareCaptionJustification{justified}{\justifying}
\usepackage[]{algorithm}
\makeatletter
\renewcommand{\ALG@name}{Function}
\makeatother

\usepackage{algpseudocode}

\usepackage{booktabs}

\renewcommand{\theenumi}{\roman{enumi}}%
\setenumerate{noitemsep}
\setitemize{noitemsep}

\usepackage[capitalise]{cleveref}

\newtheorem{definition}{Definition}

\newtheorem{lemma}[definition]{Lemma}

\begin{document}

\title{Towards interpretable quantum machine learning via single-photon quantum walks}

\author{Fulvio Flamini}
\thanks{These authors contributed equally to this work.\\ marius.krumm@uibk.ac.at}
\author{Marius Krumm}
\thanks{These authors contributed equally to this work.\\ marius.krumm@uibk.ac.at}
\author{Lukas J. Fiderer}
\affiliation{Universit\"{a}t Innsbruck, Institut f\"{u}r Theoretische Physik, Technikerstraße 21a, A-6020 Innsbruck, Austria}

\author{Thomas M\"{u}ller}
\affiliation{Department of Philosophy, University of Konstanz, Universit\"{a}tsstraße 10, 78464 Konstanz, Germany}

\author{Hans J. Briegel}
\affiliation{Universit\"{a}t Innsbruck, Institut f\"{u}r Theoretische Physik, Technikerstraße 21a, A-6020 Innsbruck, Austria}

\begin{abstract}
Variational quantum algorithms represent a promising approach to quantum machine learning where classical neural networks are replaced by parametrized quantum circuits. However, both approaches suffer from a clear limitation, that is a lack of interpretability.
Here, we present a variational method to quantize projective simulation (PS), a reinforcement learning model aimed at interpretable artificial intelligence. Decision making in PS is modeled as a random walk on a graph describing the agent’s memory. To implement the quantized model, we consider quantum walks of single photons in a lattice of tunable Mach-Zehnder interferometers trained via variational algorithms. Using an example from transfer learning, we show that the quantized PS model can exploit quantum interference to acquire capabilities beyond those of its classical counterpart.
Finally, we discuss the role of quantum interference for training and tracing the decision making process, paving the way for realizations of interpretable quantum learning agents.
\end{abstract}

\maketitle

\section{Introduction}

Classical machine learning methods are revolutionizing science and technology, with applications ranging from drug discovery \cite{Jumper2021} to the study of quantum phases of matter \cite{Venderley18}. Consequently, the exploitation of quantum effects to enhance machine learning has emerged as a rapidly evolving research domain \cite{QML}, showcasing early successes such as provable learning advantages \cite{QMLadvantage} and the development of variational quantum circuits tailored to noisy intermediate-scale quantum devices \cite{VQC}. These circuits can serve as replacements for classical neural networks within traditional algorithms. Nevertheless, amidst the rapid progress made in both classical and quantum machine learning, it is essential to bear in mind the inherent limitations of classical neural networks, which may persist or even be exacerbated in their quantum counterparts.

In particular, the opaque nature of classical neural networks can prove challenging, especially in scenarios necessitating delicate decision-making directly impacting human lives. Similarly, the quest for deeper comprehension of natural phenomena extends beyond a mere oracle for specific problems: Often the process leading to a solution yields insights that are as valuable as the solution itself.
While various approaches have been explored in the machine learning literature \cite{XAI}, the most principled approach appears to be the design of transparent models that enable the tracing of decision processes. However, with the inclusion of quantum effects, the prospect of interpreting decision processes appears daunting.
Interesting first steps in this direction, using Shapley values to measure feature importance \cite{InterpretableVQC1} or mutual information to trace the flow of information \cite{InterpretableVQC2}, still rely on black-box methods from classical explainability research.

In this work, we aim to dispel such reservations by quantizing a transparent classical reinforcement learning model known as projective simulation (PS) \cite{Briegel12, Mautner2015}. By undertaking this quantization, we lay the groundwork for investigating the interpretability aspects inherent in quantum machine learning models.
In PS, the agent's decision making process is modeled as a random walk of an excitation in an episodic memory. Following an approach already investigated by some of the authors \cite{Flamini20}, we replace this classical random walk with a single-photon quantum walk in an optical interferometer. On the one hand, we are able to preserve the core features of classical PS, so that the main ideas developed in the field can be transferred to the quantum domain. On the other hand, this approach can be tailored to end applications, and even extended to other platforms such as trapped ions \cite{Friis15, Sriarunothai19} and superconducting quantum circuits \cite{Lamata17, Cardenas-Lopez18}. These two points link to another main motivation of this work, that is, the development of an interpretable learning agent that operates in distributed quantum optical networks \cite{Wei22}, and that can interact with other agents by exchanging quantum information encoded in light.

The article is structured as follows: In Sec.~\ref{Section:ReinforcementLearning}, we review the PS model that provides the background of our work. In Sec.~\ref{Section:OverviewQPS}, we present our approach to quantizing PS and, based on that, to designing a photonic implementation. In Sec.~\ref{Section:Training}, we introduce two variational training algorithms tailored to the proposed framework,
with applications that go even beyond the scope of this work.
In Sec.~\ref{Section:TransferLearning}, we provide numerical evidence that such quantum PS agents can generalize better than their classical counterparts.
Most of our analyses can be tested with a package we make available online \cite{Package}, and can be further extended in various directions (using the same package; see also Apps.~A-C).


\section{Projective simulation}
\label{Section:ReinforcementLearning}

\begin{figure}[t]
    \includegraphics[width = 0.87 \linewidth]{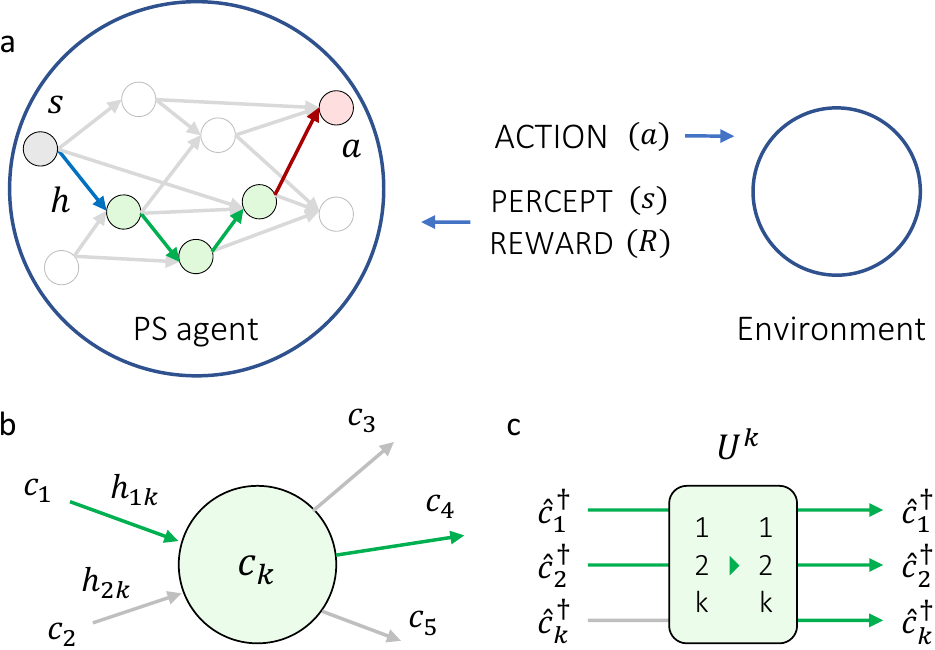}
    \caption{\textbf{Projective Simulation (PS).}
    This work uses PS as a framework for reinforcement learning and its quantization. a) When a PS agent receives a percept $s$ from the environment, it initiates a random walk in the graph that describes its memory, until it hits a vertex representing an action $a$. The performed action may be rewarded, leading to an update of the parameters ($h$-values) that control its decision making process.
    b) In classical PS, the vertices of the graph represent \emph{clips} ($c_k$). Directed edges are associated with transition probabilities that depend on the $h$-values.
    c) In the quantized PS, clips are represented by mode operators $\hat{c}^{\dagger}_k$ describing a single excitation, while the connectivity is governed by mode-mixing transformations $U^k$. Each $U^k$ acts on the mode operators associated with the parent clips of $c_k$ (here $c_1$, $c_2$), and its complex elements are the transition amplitudes that replace the classical probabilities of transitioning to $c_k$. All panels: all edges that contributed to the decision making process are in green.
    }
    \label{fig:1}
\end{figure}

In this section, we review the basics of RL and PS. RL is a paradigm of machine learning that involves an agent interacting with an environment (see Fig.~\ref{fig:1}). The agent receives from the environment an input $s \in S$, called \emph{state} or \emph{percept}, and decides what action $a \in A$ to take next. RL typically considers Markov decision problems, for which the agent's \emph{policy} can be described by a conditional probability distribution $\pi(a|s)$.
The agent receives immediate feedback on its actions through a reward $R$; its goal is to find the policy that maximizes the long-term expected reward.

\begin{figure}[t!]
    \includegraphics[width = 0.7 \linewidth]{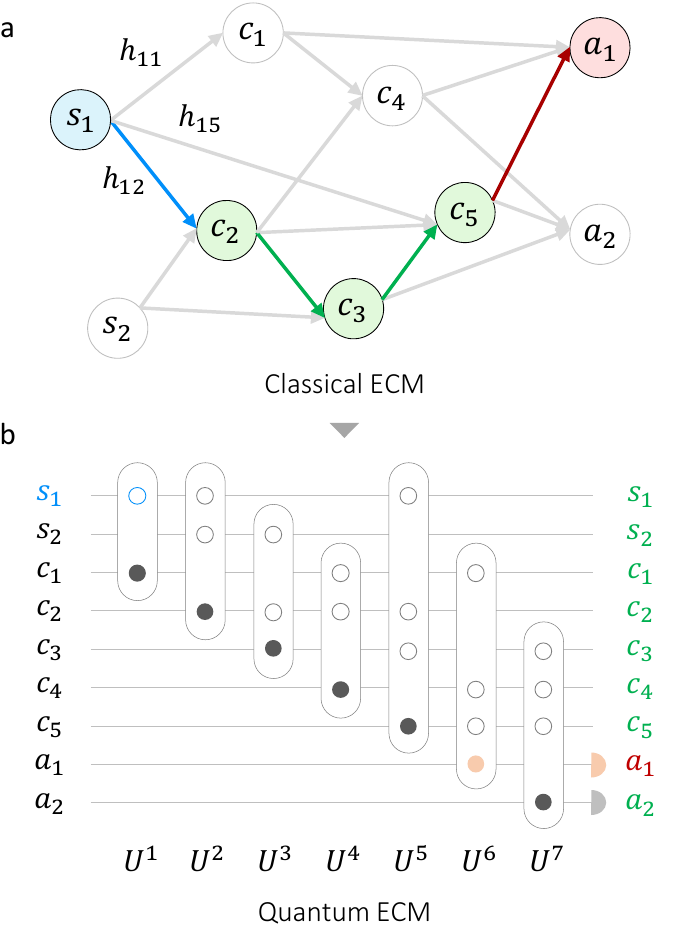}
    \caption{\textbf{Decision making in classical and quantum projective simulation (PS).} a) In PS, decision making occurs via a classical random walk over a network of clips $\{c_1, \dotsc, c_{|C|}\}$ that describes the agent's episodic and compositional memory (ECM) \cite{PSisnotNN}. When the agent receives a percept $s$ (blue), an excitation triggers and performs a random walk over the network of intermediate clips $\{c_1, \dotsc, c_{|I|}\}$ (green) until it hits one of the action clips $\{a_1, \dotsc, a_{|A|}\}$ (red), which is immediately coupled out.
    b) In the proposed quantized PS, these random walks become quantum walks of single excitations through layers implementing the transformations $U^k$, which mix sets of clips in a certain order (see App.~E). Here, clip-to-clip transition probabilities, classically described by the $h$-values, are replaced by transition amplitudes described by the matrix elements of $U^k$.
    The evolution can be seen as hopping between eigenstates (of the modes) in the Hilbert space (empty disks: in/out modes; filled disks: output modes). This panel reproduces the quantized evolution in the classical ECM in panel (a). Clips are now visited in superposition (green).
    }
    \label{fig:2}
\end{figure}

In this work we focus on PS, a physically motivated RL model that has been applied to the study of collective behavior \cite{Lopez-Incera20}, robotics \cite{Hangl16}, automated design of quantum experiments \cite{Melnikov18} and protocols \cite{Wallnofer20}, and quantum error correction \cite{PoulsenNautrup19}.
PS relies on a memory structure called \emph{episodic and compositional memory} (ECM),  described by a directed graph (see Fig.~\ref{fig:1}).
The vertices of this graph, called \emph{clips}, are associated with memories of the agent such as remembered percepts, actions or intermediate, more complex steps in the decision making process. Importantly, this means that a clip carries semantic information \cite{PSisnotNN}. When the environment delivers a percept $s$, the corresponding clip is excited, and a random walk of a single excitation is initiated in its clip network (see Fig.~\ref{fig:2}), until the excitation hits an action clip $a$. This random walk in the ECM can be interpreted as a chain-of-thought that led to the decision.
Each directed edge from clip $c_i$ to clip $c_j$ is associated with an unnormalized transition probability $h_{ij}$ called $h$-value. The transition probability $p(c_j \vert c_i)$ is computed by normalizing over all clips that can be reached from $c_i$:
\begin{equation}
	p_{ij} \equiv  p(c_j \vert c_i) = \frac{h_{i j}}{\sum\limits_{j'} h_{i j'}} .
 \label{Equation:ClipTransitionProbability}
\end{equation} 
Learning occurs by changing the topology or the edge weights of the clip network, i.e. by tuning the $h$-values, which are initially set to 1. To this end, each edge with $h$-value $h_{ij}$ is additionally equipped with a so-called glow factor $g_{ij}$, which contributes to the standard PS update rule \cite{ Mautner2015}: 
\begin{align}
	h^{\mathrm{(t+1)}}_{ij} = 1+ (1 -\gamma) (h^{\mathrm{(t)}}_{ij} -1) + g^{\mathrm{(t+1)}}_{ij} R .
	\label{Equation:PSupdate}
\end{align}
Here, the role of the \emph{glow} factor $g_{ij}$ is to remember transitions ($c_i \rightarrow c_j$) taken in past interactions with the environment, so that they are reinforced when a reward $R$ is received. Further, $\gamma \in [0, 1]$ is a \textit{forgetting} factor: its main purpose is to gradually dampen previously learned transition probabilities, to let the agent adapt to changing environments. $\gamma$ is also used as a soft cut-off to prevent the $h$-values to overflow. Each $g_{ij}$ is updated as:
\begin{align}
    g^{\mathrm{(t+1)}}_{ij} = \begin{cases}
        1 & \text {if } c_i \rightarrow c_j \text{ was taken at } t,\\
        (1-\eta) \ g^{\mathrm{(t)}}_{ij} & \text{ otherwise,} \end{cases}
    \label{Equation:Glow}
\end{align}
where $g^{\mathrm{(0)}}_{ij}=0$ and $\eta \in [0, 1]$ is a discount factor for the actions taken further in the past. We refer to the literature on PS for a more detailed description and additional mechanisms, such as association \cite{Melnikov17} and reflection \cite{Paparo14, Dunjko15}.

\section{Quantum projective simulation}
\label{Section:OverviewQPS}

Since decision making in PS is generated via a random walk of an excitation, it is possible to quantize it by giving learning agents access to a quantum walk in a quantum  memory. In the literature, several variants of this idea have been investigated. In Ref.~\cite{Paparo14}, an algorithm based on discrete quantum walks was presented which provides a quadratic speedup. However, this algorithm relies on sophisticated oracles that likely will not be available in the near future. Furthermore, since the algorithm is described as a quantum circuit, it is not clear how it can be interpreted as a random walk of a physical excitation. On the other hand, in Ref.~\cite{Briegel12}, it was proposed to quantize PS by considering a quantum walk of an excitation in a quantum many-body system. Since Hamiltonians are hermitian, in such systems the probability amplitude of a transition is as large as the amplitude of the backwards transition. To break this symmetry, the authors proposed to use dissipative effects (`quantum jumps'), at the cost of coherence.

In this section, we propose a quantum extension of PS that allows for an optical implementation. Our guiding principle is to design a PS agent \emph{natively} as a quantum walk of a physical excitation. In particular, our proposal does not rely on the availability of quantum oracles. Our scheme uses an optical interferometer that induces a natural directionality of transitions, without the need for dissipation. We will first describe how to implement the main elements of PS in the quantum domain in general and then, in Sec.~\ref{Section:OpticalPS}, on an optical platform.


\subsection{From classical to quantum projective simulation}
\label{Section:fromPStoQPS}

In order to quantize PS and facilitate an optical implementation, we need to quantize its main elements: the episodic memory, realized by the clip network, and the decision-making process, described by the propagation of an excitation through the clip network. A physical realization of a clip network could be a system of harmonic oscillators (modes) with non-trivial many-body interactions between sets of oscillators \cite{Briegel12}. In case of an optical implementation, clips $c_i \in C$ (including percept clips, action clips and intermediate clips) then correspond to modes in an optical interferometric network \cite{Flamini20} (see Sec.~\ref{Section:OpticalPS}). For the quantization, we associate with each clip $c_i$ a pair of creation and annihilation operators of the corresponding (optical) mode, $\{\hat{c}^{\dagger}_i, \hat{c}_i\}$. An excitation of memory clip $c_i$ at time $\tau_0$ is then described by the quantum state $\hat{c}^{\dagger}_i (\tau_0) \ket{\mathrm{vac}}$, where $\ket{\mathrm{vac}}$ denotes the vacuum state, describing the inactive state of the agent's memory. Here we use the Heisenberg picture where the operators carry the agent's internal time dependence. Note that we index RL steps with $t$, while physical time within a step is denoted by $\tau_k$.

A decision-making process is initiated by the excitation of some clip $c_i$ at time $\tau_0$. The subsequent evolution of the agent's memory will be described by a unitary of the form
\begin{align}
    \hat{c}^\dagger_j (\tau_n )=\mathcal{U}^{\dag}(\tau_n, \tau_0) \, \hat{c}^\dagger_j(\tau_0) \, \mathcal{U}(\tau_n, \tau_0) = \sum _{i=1}^{|C|}U^{\textup{ECM}}_{ji} \ \hat{c}^\dagger_i(\tau_0) \ ,
\end{align}
which relates the excitation operators of the clip network at different times. The Heisenberg time evolution of the ECM $\mathcal{U}(\tau_n, \tau_0)$ is thus realized by the interferometric scattering matrix $U^{\textup{ECM}}$ connecting the modes, which, in turn, is composed of a sequence of $n$ intermediate transformations $U^k$,
\begin{align}
    U^{\textup{ECM}}(\vec \theta) = \prod_{k=1}^n U^k (\vec \theta) \ .
\end{align}
\noindent The $U^k$ are updated during the learning process via the set of parameters $\vec \theta$ (see Fig.~\ref{fig:2}).
The precise role of the $U^k (\vec \theta)$ will be discussed in more detail below. For convenience and to facilitate an optical implementation, in this work we will consider ECMs described by a directed acyclic graph (DAG). \\

\textit{Percept encoding \textemdash}
When the agent receives a classical percept at time $\tau_0$, this leads to the excitation of a percept clip $s$. The quantum state of the memory is then described by $c^{\dagger}_{s}(\tau_0) \ket{\mathrm{vac}}$. Subsequently, a quantum walk through the clip network starts \cite{Briegel12}. \\

\textit{Quantum walk in the ECM \textemdash} To connect the random walk in the ECM to the quantum walk considered in this work, we refer to the examples in Figs.~\ref{fig:1} and~\ref{fig:2}. As shown in panel Fig.~\ref{fig:1}b, a clip effectively acts as a node that routes incoming edges from sets of clips to other clips in the next step of the random walk. In the quantum ECM sketched in Fig.~\ref{fig:1}c and Fig.~\ref{fig:2}b, this mechanism translates into the application of a transformation $U^k$, which mixes creation operators of incoming clips with $\hat{c}_k^\dagger$.
A general clip-to-clip transition $c_i \rightarrow c_j$ is then modeled by 
transformations $U^k$ that mix subsets of creation operators, so that the stochastic nature of PS manifests itself at the level of quantum amplitudes.

A crucial departure from classical PS is that clips are now visited in parallel (in superposition, see Fig.~\ref{fig:1}c), and all regions of the ECM that are reachable from a percept via a sequence of $U^k$ contribute to the decision making process. As a consequence, the probability to find an excitation in action clip $a$ at time $\tau_n$ (described by the quantum state $c^\dagger_a(\tau_n) \ket{\mathrm{vac}}$) when starting with an excitation of percept clip $s$ at time $\tau_0$ (described by $c^\dagger_s(\tau_0)\ket{\mathrm{vac}}$) is given by the modulus square of the corresponding transition amplitude
\begin{align}
    p_{sa} = \left| \braket{a | s} \right| ^2 \equiv
     \left| \braket{a; \tau_n | s; \tau_0} \right| ^2 =
                       \left| U^{\textup{ECM}}_{as} (\vec \theta ) \right| ^2 , \label{eq:psa}
\end{align}
\noindent where $\ket{c_i;\tau_k} = \hat c_{i}^\dagger (\tau_k)\ket{\mathrm{vac}}$ are the time-dependent eigenstates of the number operators $\hat c_{i}^\dagger (\tau_k)\hat c_{i} (\tau_k)$ corresponding to an excitation being present in clip $c_i$ at time $\tau_k$. Eq.~\eqref{eq:psa}  entails a quantum interference of all paths. While in classical PS the excitation has a well-defined trajectory in the ECM, now the excitation can be delocalized over all the allowed modes (i.e., the quantum amplitude is in general non-zero for more than one mode at the same time). This aspect bears an interesting potential for the learning model, which has been discussed in the literature \cite{Briegel12, Saggio21} and that is further discussed in Sec.~\ref{Section:Training}.\\

\textit{Action decoding \textemdash} The decision-making process ends when $U^n$ is applied and the output states are measured. When the excitation is detected in an output state associated with an action, this action is coupled out.
In all other cases, the excitation can be (i) routed back to other regions of the reachable ECM; (ii) discarded, exciting again the same percept. Solution (ii) can be interpreted as the agent's capacity to \textit{not} take action during this decision-making process (see also the notion of \textit{suspension of judgment} \cite{Wagner22}).

Practically, the agent takes action by post-selecting over the outcomes that correspond to actions. In general, one can associate multiple output states with a single action.
This makes it possible to avoid post-selection in settings with an action space smaller than the percept space, or to implement quantum channels that are more general than unitaries.\\


\textit{Training the quantum PS agent \textemdash}
To train the agent, we consider a variational approach based on a loss function $\mathcal L$ that is inspired by classical PS and aims to reproduce its learning mechanism. $\mathcal L$ depends on the probabilities $p_{s a}(\vec \theta)$ to take action $a$ given a percept $s$ (that is, the policy) and the variational parameters $\vec \theta$. These probabilities can be estimated either numerically (by simulating the quantum walk) or experimentally (via multiple measurements with fixed input). In Sec.~\ref{Section:LossFunction} we introduce and motivate our choice for $\mathcal L$ (see Eq.~\ref{Equation:QPSloss}), along with several modifications and training methods.


\subsection{Quantum optical projective simulation}
\label{Section:OpticalPS}

In this section, we describe our approach to implement quantum PS on a photonic platform.
Table \ref{table:C2QmapPS} summarizes the main connections with classical PS.\\

\begin{table*}[t]
    \begin{center} 
        \begin{tabular}{p{0.21\linewidth}p{0.27\linewidth}p{0.44\linewidth}} 
        Element    & Classical PS     & Quantum PS \: ($\rightarrow$ Optical implementation)       \\ 
        \midrule
        Decision making    & Random walk of one excitation   & Quantum walk of a single excitation (photon) \\
        Agent's memory   & Directed graph of clips    & Variational (optical) circuit \\
        Clips       & Vertices of the graph     & (Optical) mode operators \\
        Clip-to-clip transitions      & $h$-values     & Quantum scattering amplitudes    \\ 
        Input/Output encoding   & Percept/Action clips   & Input/Output modes \\
        Learning mechanism & Update of $h$-values via Eq.~\eqref{Equation:PSupdate} & Update of variational parameters (phase shifters) via Eq.~\eqref{Equation:QPSloss}\\
        \bottomrule
    \end{tabular}
    \caption{Links between the main elements of projective simulation (PS) and its proposed optical quantization.}
    \label{table:C2QmapPS}
    \end{center}
\end{table*}

\textit{Related works \textemdash} Photonic circuits provide a promising platform in both classical and quantum domains \cite{Wang20,Bogaerts20}, with results of increasing complexity being continuously reported \cite{Taballione22, Arrazola21}, in particular for (un)supervised learning \cite{Shen17, Steinbrecher19, Williamson19, Marquez21, Zhang21, Xu21} and RL \cite{Flamini20, Saggio21, Lamata21} based on path encoding.

In this work, we focus on RL and aim to design a quantum PS agent that operates with a native quantum walk of physical excitations. Hence, we will replace the single excitation with a single photon propagating in a linear-optical circuit. This approach, while retracing some of the intuitions in Ref. \cite{Flamini20} where the $p_{sa}$ were reproduced by optical binary trees, promises the following advantages: (1) it requires only a single (possibly larger) circuit for all percepts, instead of one binary tree per percept and layer, with no explicit need for fast active switching; (2) it allows to take advantage of quantum coherence and interference in the decision making process; (3) it allows for more advanced training algorithms, which are applicable beyond the scope of this work (see Sec.~\ref{Section:Training}); (4) it suggests concrete directions for further developments (see Apps.~B-C), such as a framework for learning with quantum data. Below we outline how the quantized version of PS can be realized on a photonic circuit.\\

\begin{figure}[t]
	\includegraphics[width=0.8\linewidth]{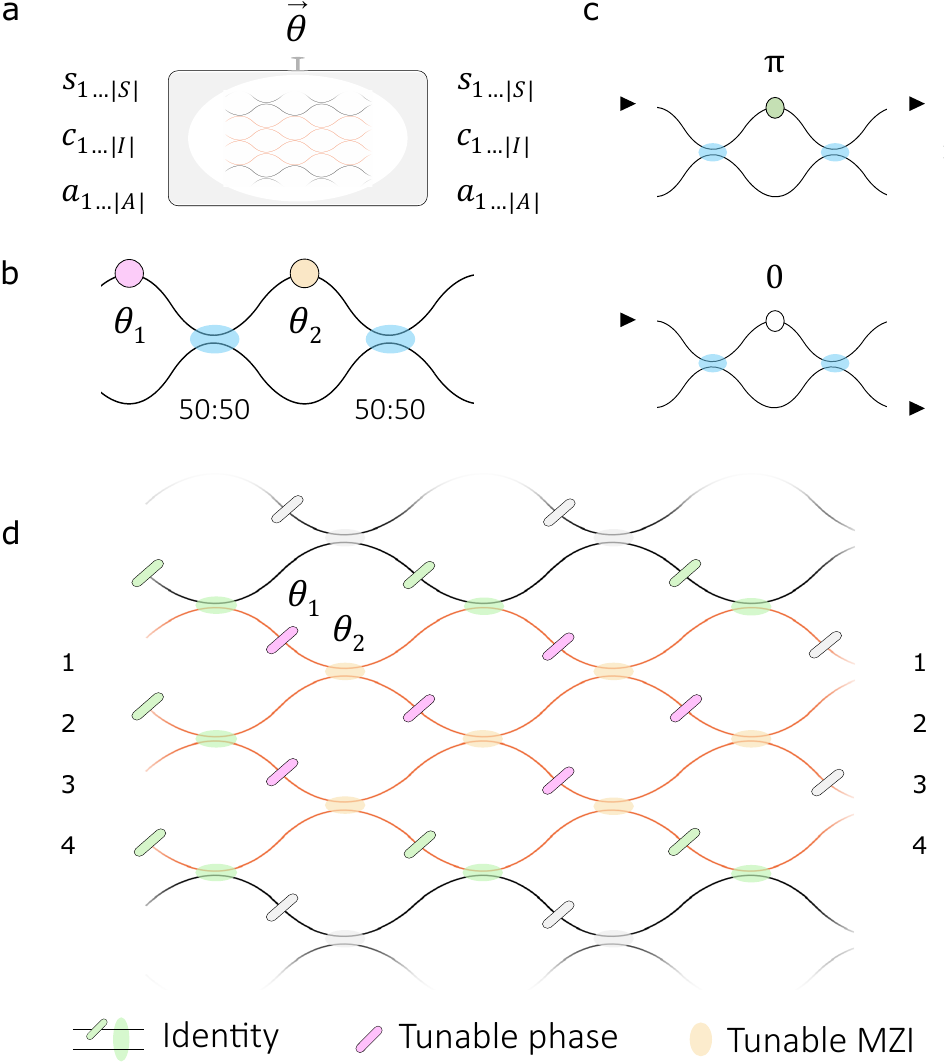}
	\caption{\textbf{Optical implementation of PS.} To enable PS on photonic circuits, we use as variational ansatz the square architecture \cite{Clements16}. a) The update of the agent's memory translates into an update of the phase shifters $\vec{\theta}$. b) Any unitary $U$ can be implemented with a regular mesh of Mach-Zehnder interferometers (MZI), consisting of two 50:50 beam splitters (BS, blue) and two phase shifters ($\theta_1$, $\theta_2$) \cite{Reck94, Clements16}. c) When $\theta =\pi$ all light entering a MZI is reflected (\textit{bar} state), while when $\theta =0$ it is transmitted (\textit{cross} state).
	d) Square, universal architectures can be carved inside larger ones by tuning the phases around them. In this panel, a universal $4 \times 4$ circuit (in/out modes are labeled) is carved in a larger mesh. Tilted elements represent tunable phase shifters (those inside the carved structure are in fuchsia). BSs with arbitrary transmissivities (orange) are implemented using tunable MZIs. Light can be confined to the pink waveguides by setting the green optical elements.
	}
	\label{fig:optics}
\end{figure} 

\textit{Photonic architecture \textemdash}
For the implementation of PS on quantum hardware, we want to identify quantum extensions of the Markov chain process that govern the evolution of a classical excitation. These extensions, usually referred to as quantum channels (completely positive trace-preserving maps), would have to reproduce the dynamics in classical PS as a limiting case.
 A direct implementation of classical PS would require the $\{ U^k \}$ to reproduce the stochastic maps that form the random walk in the ECM graph described in Sec.~\ref{Section:ReinforcementLearning}. Specifically, the real-valued matrices $P$ associated with these transformations need to (i) reproduce the probabilities $p_{ij}$ induced by the $h$-values, and (ii) fulfill the connectivity in the ECM ($P$ can be seen as a weighted adjacency matrix). Such a class of transformations can in principle be realized on a photonic platform by recalling that any real-valued matrix $P$ can be decomposed as $P=V_1 \Sigma V_2^{\dagger}$ via singular value decomposition, where $\Sigma$ is a diagonal matrix with non-negative real entries (i.e., optical attenuators) and $V_{1,2}$ are unitary transformations \cite{Shen17}. Hence, the architecture we use for the variational circuit is the canonical square one \cite{Clements16}, since it can implement any unitary transformation by suitably tuning its phase shifters. Indeed, we recall that any such $U^k$ can be implemented by a homogeneous mesh of Mach-Zehnder interferometers, each comprising two 50:50 beam splitters and two phase shifters (see Fig.~\ref{fig:optics} and Sec.~\ref{Section:SM_Ansatz}) \cite{Reck94, Clements16}.
The variational parameters $\vec \theta$ correspond to the phase settings that control $U^k$.

In the fully coherent case, without additional attenuators, one can set $\Sigma = I$ in the singular value decomposition, meaning that the $U^k$ become unitary. In this case, due to unitarity constraints and since each $U^k$ acts as an adjacency matrix, additional edges between clips are created that are not present in the original ECM. This limitation represents an additional departure from classical PS, which can open up interesting opportunities in the context of learning, for instance leveraging the constraints to improve generalization.\\

\textit{Percept encoding \textemdash}
We associate each percept with a photon input in a spatial mode of the circuit. The size of the Hilbert space grows linearly in the number of percepts.\\

\textit{ECM structure \textemdash} We associate each clip with one optical mode of the architecture. The unnormalized transition probabilities $h_{ij}$ that describe the transitions $c_i \rightarrow c_j$ are now replaced by transition amplitudes between optical modes, described by a complex amplitude $U_{ji}^k$.

Our chosen architecture is especially useful for reasons that go beyond its universality, noise resilience and compactness \cite{Clements16}, since it also promises to support key features of PS:

\begin{itemize}
    \item \textit{Clip creation/deletion}: new clips (i.e., optical modes) can be added during the learning process, while clips that are rarely traversed can be removed.
    \item \textit{Edge creation/deletion}: the random walk can be manipulated by creating or deleting links between clips.
    \item \textit{Clip composition}: fictitious clips can also be created, by random variation or merging of existing clips.
\end{itemize}

\noindent We observe that this mesh allows one to \textit{carve} substructures that are in turn universal for unitary transformations, by adjusting the surrounding phases (without the need to modify the hardware; see Fig.~\ref{fig:optics}d). This means that the Hilbert space describing the ECM can be dynamically enlarged to accommodate new clips, and the $U_{ji}^k$ can be controlled by intervening on the regions of the circuit that implement $U^k$.\\

\textit{Action decoding \textemdash} In PS, the decision-making process ends when an action clip is hit. In the proposed implementation, this corresponds to detecting photons from one of the output modes that correspond to an action clip. Whenever a photon is detected in a mode that does \textit{not} correspond to an action, that event is discarded. In the future, these photons could be routed back into the circuit, or associated with mechanisms of suspension of belief \cite{Wagner22}.\\


\section{Training the architecture}
\label{Section:Training}

In this section, we discuss two algorithms to train the quantum optical PS agent. To this end, we recall that the standard update rule in PS has no straightforward quantum analogue: while only one path is taken by the classical excitation and rewarded by the update rule (technically, the $h$-values associated with its edges), all paths from percept to action are visited by the quantum excitation, and the very notion of path loses its meaning (see App.~D for a discussion on the impact of superposition in the quantized model). Hence, also motivated by the advances in quantum machine learning, here we take a different, variational approach to update the parameters $\vec{\theta}$.


\subsection{Variational approach to mimic classical PS}
\label{Section:LossFunction}

The first algorithm to train the quantum PS agent is based on a variational optimization of a loss function $\mathcal{L}$ inspired by classical standard PS.

Since a quantum circuit does not have access to the intermediate states of the computation, the update rule we propose adapts that of a classical 2-layer ECM. This does not mean that the quantum agent is necessarily 2-layered (in fact, the architecture supports multi-layer ECMs, as shown in Fig.~\ref{fig:2}); rather, it means that we update the observable transition probabilities $p_{sa} (\vec \theta)$. In turn, also the intermediate transition amplitudes in the circuit are affected. However, the update rule of PS requires the use of unnormalized $h$-values instead of (normalized) probabilities. Thus, let us first introduce the classical normalization factor of the $h$-values $h_s$, i.e. $h_s := \sum_{a} h_{sa}$, from which we get $h_{sa} = p_{sa} h_s$ and the classical 2-layer PS update rule
\begin{align}
    p_{sa}({\vec \theta}^{\mathrm{(t+1)}})  h^{\mathrm{(t+1)}}_s =
    1+ (1-\gamma) \left(p^{\mathrm{(t)}}_{sa} h_s^{\mathrm{(t)}} -1\right) + g^{\mathrm{(t)}}_{sa} R . \label{Equation:ExactUpdate2Layer}
\end{align}
A loss function $\mathcal{L}$ that implements Eq.~\eqref{Equation:ExactUpdate2Layer} is
\begin{widetext}
    \begin{equation}
        \mathcal{L}({\vec \theta}^{\mathrm{(t+1)}}, \{ h_s^{\mathrm{(t+1)}} \}_s ) =
        \sum_{s} D \Big[ p_{sa}(\vec \theta^{\mathrm{(t+1)}}) h^{\mathrm{(t+1)}}_s - 1 ,
        \ (1-\gamma) \left(p^{\mathrm{(t)}}_{sa} {h^{\mathrm{(t)}}_s} -1 \right) + g^{\mathrm{(t)}}_{sa}  R  \Big]
        \label{Equation:QPSloss}
     \end{equation}
\end{widetext}
\noindent Here, $\vec \theta^{(t+1)}$ and $\vec h_s^{(t+1)}$ are variational parameters, while $D$ is a suitable distance measure such as the Kullback-Leibler divergence (after rearranging the arguments of $D$), $R$ is the reward, and $g_{sa}$ is the glow factor described in Eq.~\eqref{Equation:Glow}. The expression in Eq.~\eqref{Equation:Glow} is still valid in the proposed quantum extension, since the update rule underlying the loss function is taken from a classical two-layer PS, and both input ($s$) and output ($a$) are classical. However, we emphasize again that the quantum ECM itself is multi-layered.
Since uncontrolled output phases can spoil the quantum interference, if the agent is used as a module in a larger photonic setup, we choose to add the $\ell^1$-norm of these phases to the loss function.

Note that only the first argument of $D$ in Eq.~\eqref{Equation:QPSloss} contains tunable parameters, while the right entry contains the target values calculated from the previous state of the agent, as a reference, similar to deep-Q learning. Similarly, one can use \emph{experience replay} \cite{Lin1992, ExperienceReplay} to collect rewards and glows for a batch of percept-action pairs, and add up or merge the loss functions in appropriate ways. That is because continuous online learning (i.e. updating after each action) has the disadvantage that the unitarity constraints will also affect the transition probabilities of undetected percept-action pairs, without taking their rewards into account.

In the following, we discuss suitable heuristic simplifications of this loss function, which make it more amenable to an actual implementation. First, by approximating $h_s^{\mathrm{(t)}} \approx h_s$ and introducing an effective reward $r_s := \frac{R}{h_s}$ in the exact update rule in Eq.~\eqref{Equation:ExactUpdate2Layer}, we obtain
\begin{align}
    p^{\mathrm{(t+1)}}_{sa}(\vec \theta) - \frac{1}{h_s} \approx
    (1-\gamma)  \left(p^{\mathrm{(t)}}_{sa} -\frac{1}{h_s}\right) + g^{\mathrm{(t+1})}_{sa}  r_s .
\end{align}
Without a reward (i.e. $r_s = 0$), the forgetting mechanism after the approximation seeks to reset $p^{\mathrm{(t)}}_{sa}$ to $\frac{1}{h_s}$. However, in information theory, the no-information distribution is the uniform distribution. This motivates us to replace $\frac{1}{h_s}$ with a uniform distribution $\frac{1}{|A|}$,
\begin{align}
    p^{\mathrm{(t+1)}}_{sa}(\vec \theta) =
    \frac{1}{|A|} + (1-\gamma)  \left(p^{\mathrm{(t)}}_{sa} -\frac{1}{|A|} \right) + g^{\mathrm{(t+1})}_{sa}  r_s . \label{Equation:SimplifiedUpdateRule}
\end{align}
So far, we did not make any choices concerning the source of the rewards $R$. For our simplifications, we now  take the point of view that the reward $R$ is designed by a programmer to achieve an intended agent behavior. Then, a further simplification can be made by directly modelling the effective reward $r$, rather than $R$. Therefore, we replace $r_s$ with $r$. With this change in mind, Eq.~\eqref{Equation:SimplifiedUpdateRule} now looks like an analogue of the exact update rule in Eq.~\eqref{Equation:ExactUpdate2Layer}, but for probabilities instead of $h$-values.
To ensure that the agent updates get smaller over time, it is recommended to use an ``annealing schedule'' for the reward, i.e. to put an additional time dependence into the reward of the form $r := f(t) \cdot r_0$. Here, $f(t)$ is a function that monotonically decreases from $1$ to $0$. This allows the training algorithm to make finer and finer updates to the probabilities.

In order to stay as close as possible to classical PS, the update rule would have to be applied to all percept-action pairs. This would require implementing the forgetting mechanism ($\gamma > 0$) for unvisited percept-action pairs, or enforcing the transition probability of unvisited pairs to remain unperturbed ($\gamma = 0$), leading to a complicated loss function.
To avoid this overhead, we include only the actually observed percept-action pairs in the loss function, while the transition probabilities of unobserved pairs are allowed to change according to the unitarity constraints.  Combining all of the above simplifications, we obtain the loss function
\begin{align}
	\mathcal{L}(\vec \theta ) = & \sum_{(s,a) \, | \, g^{(t)}_{sa} > 0 }D\Bigg[  p_{sa}({\vec \theta}^{\mathrm{(t+1)}} ) \ , \label{Equation:SimplifiedLoss} \\
	& C\left( \frac{1}{|A|} + (1-\gamma) \Big(p^{\mathrm{(t)}}_{sa} - \frac{1}{|A|}\Big) + g^{\mathrm{(t+1})}_{sa} r \right)\Bigg]  \nonumber 
\end{align}
\noindent where $C$ is a (smooth) cutoff function to enforce that probabilities are in $[0,1]$, and only $(s,a)$-pairs with non-zero glow are considered. 
We point out that summing over all states is necessary only in the general case, due to the forgetting and glow mechanisms. However, when the environment is described by a Markov decision process (with no forgetting) or the glow has a finite horizon, the number of states in the sum gets significantly reduced and the process faster.


\subsection{Variational approach based on causal diamonds}
\label{Section:CD}

In the previous section, we implicitly assume that at each step one tunes \textit{all} phase shifters (i.e., parameters $\vec \theta$) in the circuit.
Indeed, this is possible and can be beneficial to train the agent faster and with a better performance. At the same time, updating all phases at each step might not be necessary for the following reason.

In a square architecture with $|C|$ input/output modes, the number of phase shifters is quadratic in $|C|$. However, for a fixed transition probability $p_{sa}(\vec \theta)$, one can see that only some phases play a role: Those at the intersection of the future and past light cones of $s$ and $a$, respectively (see Fig.~\ref{fig:CausalDiamond}a). In the literature on relativity, this intersection is called \emph{causal diamond}. A number of considerations follow:

\begin{figure*}[t]
    \includegraphics[width =  \textwidth]{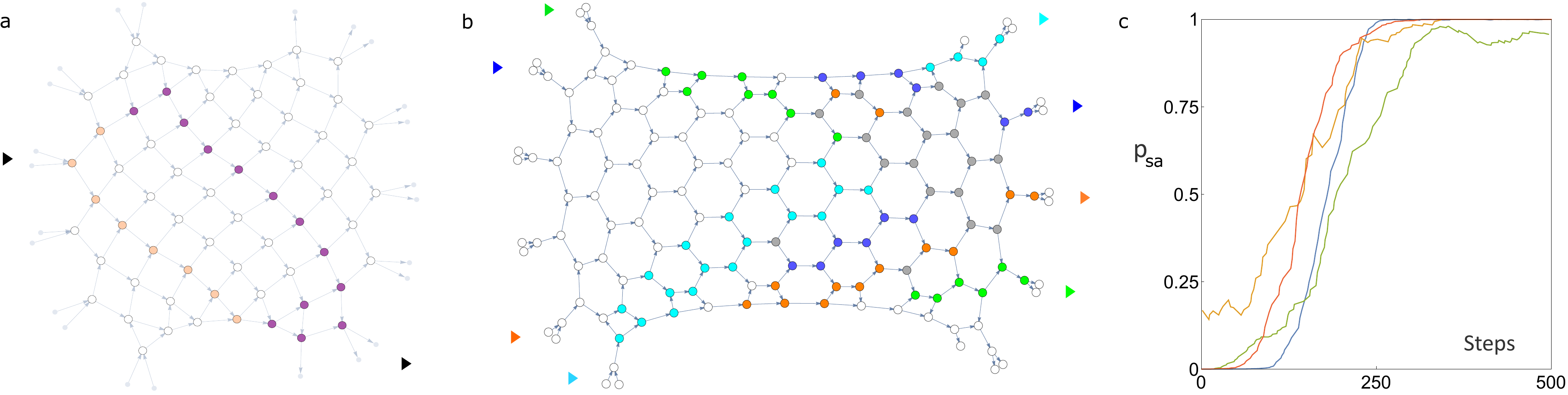}
    \caption{ \textbf{Update via causal diamonds.} To reinforce a transition probability $p_{sa}$ in a 2-layer PS, it is not necessary to update all phase shifters at each step. a) Only the phases inside the causal diamond of ($s$, $a$) influence $p_{sa}$ (colored nodes indicate the surface of the causal diamond). Since light can only leave the causal diamond from the purple nodes (\textit{leaking nodes}), it is sufficient to tune their phases. b) The approach works also with multiple transitions, by iteratively tuning different sets of leaking nodes (here shown with five different colors, including gray for the overlaps, and representing pairs of adjacent phase shifters and beam splitters as a single node), even if the updates compete with each other (some leaking nodes belong to multiple causal diamonds). c) The agent learns to maximize $p_{sa}$ for the four ($s$, $a$) pairs in panel (b), by tuning only the leaking nodes. In some cases, it may be necessary to tune more nodes to maximize all probabilities. Each percept/action is encoded in a pair of adjacent input/output modes.}
    \label{fig:CausalDiamond}
\end{figure*}

\begin{enumerate}
  \item To update $p_{sa}(\vec \theta)$, one could simply tune the phase shifters inside the causal diamond.
  \item For light to reach $a$ from $s$, what matters is that light does not leak out of their causal diamond. Hence, one can also just tune the $O(|C|)$ phases that lie on the \textit{surface} of the causal diamond.
  \item Since leakage only occurs on the surface of the past light cone, it is sufficient to focus on its intersection with the causal diamond (see Fig.~\ref{fig:CausalDiamond}). We call these phase shifters \textit{leaking nodes}.
\end{enumerate}

Overall, in terms of reconfiguration time, stability and power consumption, the above observations allow for progressively simpler experimental requirements. The price to pay is that the surfaces of different causal diamonds (for different ($s$, $a$) pairs) intersect and, for this reason and due to unitarity constraints, updating one transition probability also affects the other transitions. Also, the fewer phase shifters are tuned, the smaller the set of values that solve the optimization problem when multiple ($s$, $a$) pairs are rewarded. The desired trade-off can be found with the experimental conditions in mind.

We consider two main strategies to tune the sets of relevant phases described by points (i-iii) above: (a) phases are independently updated by gradient descent, and (b) phases are updated sequentially (from left to right, each phase in a layer is set to a new value before the phases in the next layers are adjusted). The latter strategy is possible because we assume that light propagates only forward, and it leads to a faster and smoother training. A quantitative comparison between the two strategies, and a study of the most suitable one with multiple rewarded percept/action pairs, can be done with our shared package, keeping experimental requirements in mind \cite{Package}. An illustrative analysis for strategy (b) is shown in Fig.~\ref{fig:CausalDiamond}c. Additional information and analyses can be found in Sec.~\ref{Section:SM_CD} and in App.~F, respectively.
Importantly, we point out that this approach is not restricted to the regular mesh considered in this work. Rather, similar considerations also apply to circuits with arbitrary mode connectivity and even higher-level structures in the ECM \cite{Package}.

\begin{figure}[t]
    \includegraphics[width = \linewidth]{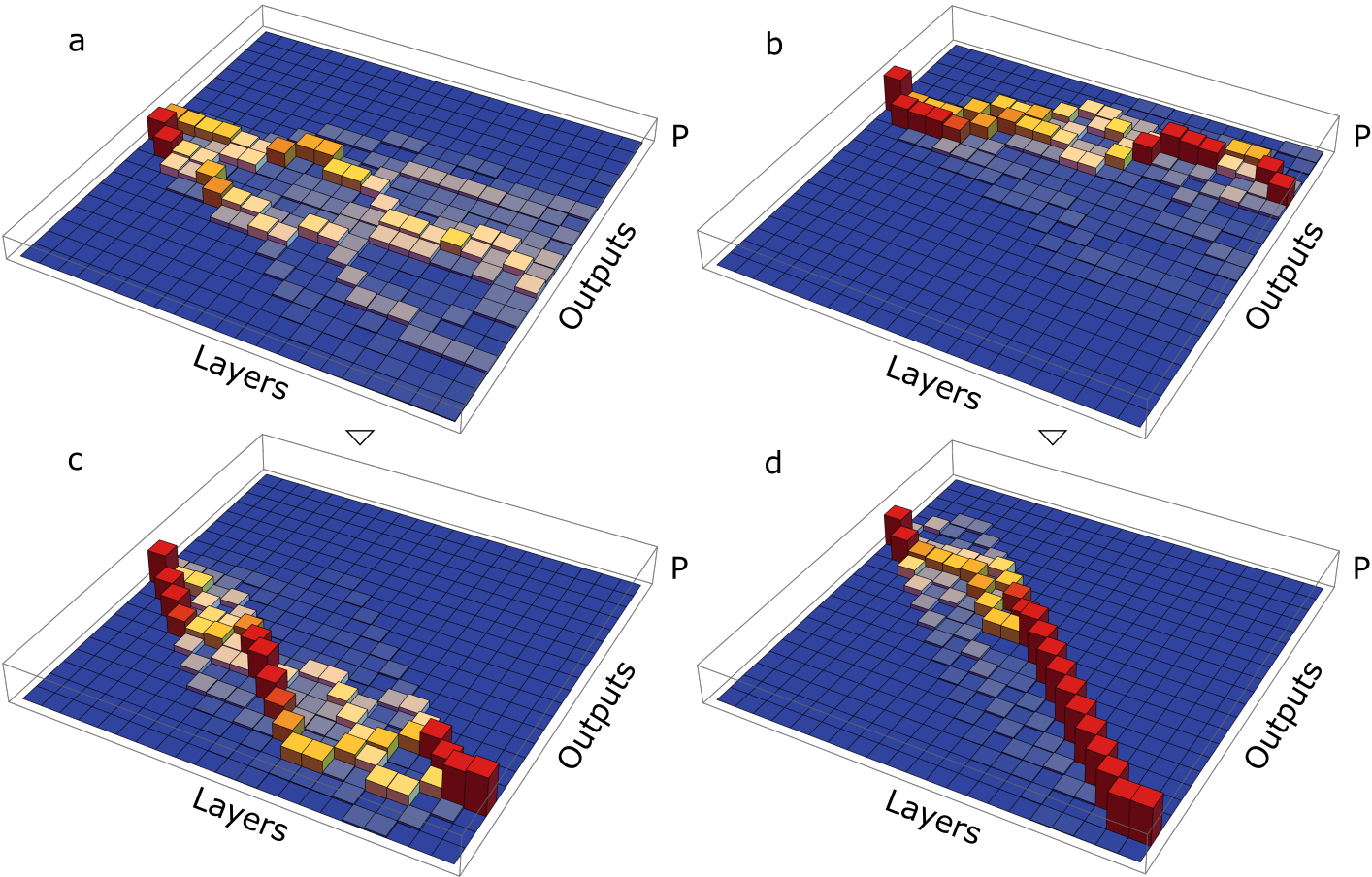}
    \caption{\textbf{Partial traceability in the decision making process of quantum PS agents.}  
    a-b) Two quantum walks, associated with as many 2-layer ECMs, are numerically simulated with a square optical architecture. Each panel shows the probability $P$ to measure a photon in one of the output modes before training. Each column (label: \textit{Layers}) displays the squared column (for a fixed percept) of the unitary matrix obtained by adding a new layer of optical elements, from left (input layer) to right (full circuit). The rightmost column of each plot corresponds to the output probabilities of a quantum walk over the full ECM. c-d) Same analysis after training with the causal diamond. Panels (c) and (d) correspond to the evolutions in panels (a) and (b), respectively. We observe that a notion of partial trace emerges inside the circuit, which can be leveraged to study interpretability in quantum learning agents.
    }
    \label{fig:5}
\end{figure}

Regarding the aspect of interpretability, we have seen how PS offers a viable framework towards achieving this goal, by allowing one to retrace the path taken between percepts and actions. We also know that this is no longer true in the quantum domain, since all paths contribute to the decision making process at once.
The causal diamond allows us to make a first step toward the interpretability of quantum PS, by observing that a notion of \textit{partial} trace can be recovered if one studies the likelihood that one photon (or more) would be measured in a given mode of the circuit (see Fig.~\ref{fig:5}). This information can be retrieved by numerically simulating the single-photon quantum walk, since all phase settings are known. This possibility still holds for multi-photon quantum walks despite the hardness of the computation, which makes it practically feasible only for a small number of photons \cite{Wu18}.

\subsection{Direct updating via a Gram-Schmidt process}
\label{Section:GSO}

The previous sections consider a variational approach to the optimization problem, where phases are gradually adjusted to produce a new transformation that behaves better under a given figure of merit. We now consider a complementary approach, where we first update the unitary evolution $U$ of the ECM, and then use a well-known decomposition \cite{Clements16} to retrieve the $\vec{\theta}$ that produce the desired $U$. In this case, one directly adjusts the matrix elements of $U$ to implement an update rule such as Eqs. \eqref{Equation:ExactUpdate2Layer} or \eqref{Equation:SimplifiedLoss}: the new matrix is no longer unitary; however, a Gram-Schmidt orthonormalization (GSO) process yields a new $U'$ that approximates the intended $U'_{a s}$.

\begin{figure}[t]
    \includegraphics[width=0.72\linewidth]{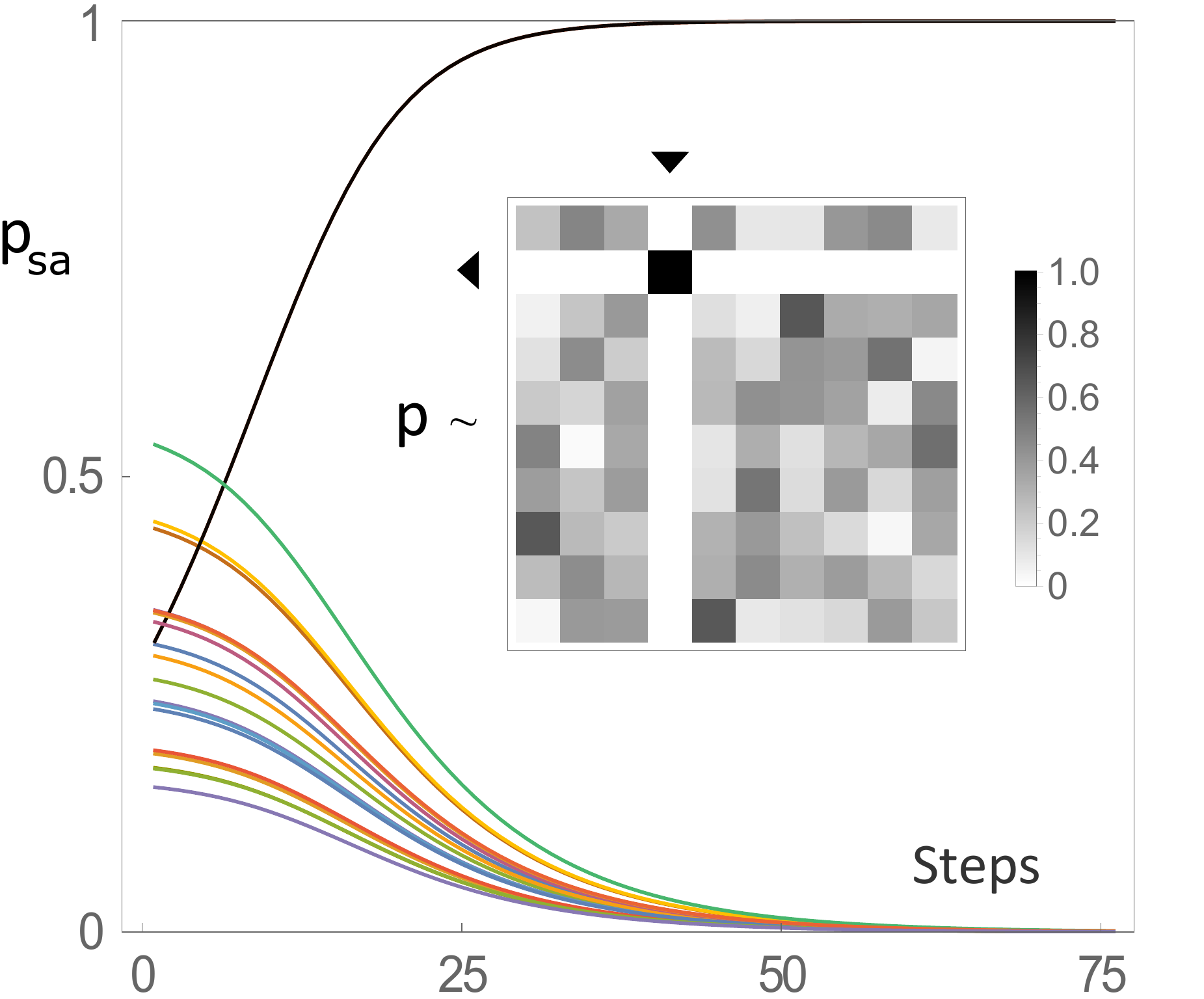}
    \caption{\textbf{Update via a Gram-Schmidt process.} We can directly update the unitary transformation $U$ that describes the ECM using a Gram-Schmidt orthonormalization. Here, a learning curve (black) is shown for a single rewarded ($s$, $a$) pair in a 2-layer PS described by a $10 \times 10$ Haar-random $U$. All other transition probabilities (other colors), from $s$ and to $a$, decrease to 0. Curves correspond to a single agent, without average. Inset: $( \lvert U_{ij} \rvert ^2 )_{ij} $ after training (arrows indicate ($s$, $a$)).}
    \label{fig:GSO}
\end{figure}

We add a few remarks on this approach (see also Sec.~\ref{Section:SM_GS}).
First, if $U'_{a s} $ is close to the old value $U_{a s} $, one step of GSO only causes a small update of the other entries of $U$.
In this sense, this approach allows for a smooth, flexible and controlled update of $U$ (see Fig.~\ref{fig:GSO}). Also, both the GSO and the unitary decomposition are efficient and relatively fast subroutines, which could even be faster than collecting statistics for other variational approaches. In other words, the overhead is outsourced to the classical CPU.
The limitation of this approach is that it assumes perfect knowledge of $U$ and a precise control of the experimental settings. The impact of imperfections can be estimated using the shared package \cite{Package}.

\section{Quantum versus classical projective simulation in a transfer-learning scenario}
\label{Section:TransferLearning}

In this section, we apply the proposed quantum PS to an actual learning problem. The PS update rule of Eqs. \eqref{Equation:ExactUpdate2Layer} and \eqref{Equation:QPSloss} has already been used in various applications~\cite{Lopez-Incera20, Hangl16, Melnikov18, PoulsenNautrup19, Wallnofer20}. Here, we consider a scenario which pronounces the differences between classical and quantum PS.
We consider a transfer-learning scenario \cite{Bozinovski2020ReminderOT}: a machine learning setting where a model trained for one task can solve another task.
The scenario we choose is that of Eva et al.~\cite{Ried19}.

In this scenario, one considers a list of particles characterized by discretized observables, and asks whether two observables have two particular values. To answer these questions, or tasks, in our adapted version of this thought experiment the agent attempts to build a meaningful representation of the particles in two stages (corresponding to the middle layer and a task layer: In stage 1, the agent learns a representation from the percept layer to the middle layer, where each node corresponds to one observable and one value of that observable.  While Eva et al. intended that this representation is learned implicitly, in our case we directly train the middle layer to achieve this representation. In stage 2, the connections from percept to middle layer are kept fixed. The task layer has two nodes, which correspond to yes/no-answers. For each question (task), we train a separate set of connections from the middle layer to the corresponding task layer.

The contrast between classical and quantum learning is rooted in the way the decision-making process handles the structured representation of the PS agents: Since the excitation of the classical PS agent is localized, it can only use knowledge about one observable and, therefore, it cannot reliably answer questions involving two observables. The quantum PS agent, however, can use interference to combine knowledge about two observables.
In our proof-of-concept analyses, described in detail in App.~D and in the shared code \cite{Package}, the quantum agent consistently achieved over $97 \%$ (weighted) accuracy for all task layers. 

Conversely, one can use analytical reasoning to see that the classical PS agent cannot achieve near perfect accuracy: Since the classical excitation is always localized in one node, in the middle layer the classical PS agent can at most know the value of one relevant observable. Therefore, the best thing the agent can do is to guess about the other observable. Since two out of three possible values of that other observable lead to a right no-answer, the best policy the classical PS agent can represent is to always answer no. Hence, the performance of the classical PS agent is upper bounded by $\frac{8}{9}$, thereby showing yet again the impact of quantum interference for learning agents.


\section{Discussion}

Significant endeavors are currently underway to leverage quantum effects for future AI applications. At the same time, it is becoming increasingly evident that there are certain use cases where interpretability in the decision-making process is crucial. In this work, we introduced a quantum learning framework obtained through the quantization of a transparent and interpretable classical learning model, aiming to explore the intricate relationship between quantum effects and interpretable decision processes. We extend the theoretical framework by proposing an implementation based on integrated photonics that is specifically tailored for quantum reinforcement learning.

Diverging from conventional, gate-based variational quantum algorithms, our approach draws inspiration not from neural networks but from the quantization of PS, an interpretable classical machine learning model. The proposed model shares the graph structure with classical PS, where decision-making is realized through a traceable and interpretable random walk. In our implementation, this random walk becomes a quantum walk of a single photon undergoing unitary transformations, and the variational parameters become the phases of a homogeneous mesh of Mach-Zehnder interferometers. In contrast to classical PS, here quantum interference effects and the constraints imposed by unitarity come into play. As we discuss, this leads to an interesting interplay between the interpretable elements inherited from classical PS and the genuine quantum effects.
We also provide evidence of the enhanced learning capabilities achieved by quantum PS. To this end we investigate a transfer learning problem where quantum PS showcases its capacity to leverage quantum interference, surpassing the capabilities of its classical counterpart.

By bridging the gap between quantum effects and interpretability, the proposed quantum learning model paves the way for future investigations and applications that benefit from the combined strengths of interpretability and quantum effects. For example, some of the ideas include (i) the use of restricted connectivity to implement domain knowledge, (ii) the search for relevant regions of the trained model exhibiting amplitude concentration (see Fig. \ref{fig:5}), and (iii) the link to the classical field of neuro-symbolic logic~\cite{NeSyReview1}, motivated by the similarity between interference effects and Boolean logic. Additional follow-up investigations regarding the photonic implementation are already outlined in the Appendix with numerical analyses. There, Apps.~II-III discuss the potential offered by multi-photon (i.e., multi-percept) and multi-frequency quantum information processing, respectively.
All results can be tested with a package we release online \cite{Package}.


\section{Methods}

\subsection{Variational algorithms}

In this section, we provide additional information on the two variational algorithms presented in Sec.~\ref{Section:Training}.

The following algorithms naturally take into account the unitarity constraint inherent to the architecture. We point out that this constraint is not necessarily a limitation and, rather, it could be advantageous to improve the agent's performance. In fact, since a machine learning ansatz with more freedom is also more prone to overfitting, it is reasonable to believe that a specialized ansatz with restrictions and symmetries might generalize better. A reduced number of degrees of freedom might also help during the training stage.


\subsubsection{Update via Gram-Schmidt process}
\label{Section:SM_GS}

Here we outline the update via a Gram-Schmidt process. The algorithm consists of four steps (i-iv):
\begin{enumerate}
  \item Compute $U^{\mathrm{(t+1)}}_{a_0 s_0}$ from the current $U^{\mathrm{(t)}}_{as}$ according to a desired update rule. Here, $s_0$ and $a_0$ are the observed percept and action of the current round, respectively. In the examples in Sec.~\ref{Section:GSO}, we considered a constant rescaling factor $\alpha$, i.e. $U^{\mathrm{(t+1)}}_{a_0 s_0} = \alpha \ U^{\mathrm{(t)}}_{a_0 s_0}$, which acts like a learning rate.
  \item Normalize the row $U^{\mathrm{(t+1)}}_{a_0 , \bullet}$.
  \item Use GSO on the other rows to obtain a new unitary, in the following way: For each row $a \ne a_0$, define
\begin{align}
    \tilde{U}^{\mathrm{(t+1)}}_{a s} := U^{\mathrm{(t)}}_{a s} - \sum_{\tilde s \in S} \sum_{\tilde{a} \in A_{a}}  \ \overline{U^{\mathrm{(t+1)}}_{\tilde{a} \tilde{s}} } \ U^{\mathrm{(t)}}_{a \tilde{s}} \ U^{\mathrm{(t+1)}}_{\tilde{a} s} ,
\end{align}
    where the overbar denotes the complex conjugation, and $A_a$ is the set of rows $\tilde{a}$ for which we have already defined $U^{\mathrm{(t+1)}}_{\tilde{a},\bullet}$, i.e. the previous $\tilde{a}$ in the loop.
    Determine $U^{\mathrm{(t+1)}}_{a , \bullet}$ from $\tilde{U}^{\mathrm{(t+1)}}_{a ,\bullet}$ via normalization.
    \item Use the decomposition \cite{Clements16} of canonical square architectures to find the phase settings that implement the new unitary. The decomposition is exact and requires a number of operations quadratic in the size of $U$.
\end{enumerate}

The main advantage of this approach is that it allows one to control any theoretical aspect of the learning process, from the update rule to the learning speed. One limitation is that it assumes perfect knowledge and control of the experimental settings. Appending to it one step of variational fine-tuning, as the one we describe in the following section, can help mitigate this issue and speed up the learning process.


\subsubsection{Update via causal diamonds}
\label{Section:SM_CD}

\renewcommand{\theenumi}{\alph{enumi}}

Here we provide additional considerations on the update based on causal diamonds. The advantage of this approach is that it is efficient (only few phase shifters are adjusted at each step) and that it is variational (it takes into account the imperfect control over the experimental settings). 

We consider two main approaches (a,b) to update a given subset $\Theta = (\theta_1, \dotsc, \theta_N)$ of $N$ phase shifters at each training step, albeit more sophisticated strategies can be devised and tailored to the hardware. These phases can correspond to the set of leaking nodes described in Sec.~\ref{Section:CD}, or to a larger set selected via other criteria.
\begin{enumerate}
    \item \textit{Gradient ascent \textemdash} Each phase $\theta \in \Theta$ is independently tuned to maximize a predefined figure of merit $\mathcal{F}$.
    \item \textit{Sequential \textemdash} Each phase $\theta \in \Theta$, in the order given by the light propagation, is tuned to independently maximize a figure of merit $\mathcal{F}$, and is immediately set to the new value before the next phase is probed.
\end{enumerate}
The latter approach (b) (see Fig.~\ref{fig:CausalDiamond}) is usually faster than (a), since each phase update starts from a progressively higher value of $\mathcal{F}$, and the individual updates within a training step do not sabotage each other. However, an update based on gradient ascent as in (a) can be beneficial to escape local minima, which are more likely to cause problems in (b). In the Appendix we provide additional considerations on this method, together with a description of the strategy used to improve the efficiency of the algorithm.


\subsection{Optical architecture}
\label{Section:SM_Ansatz}

Here we review the basics behind the operation of the optical architecture discussed in the main text. In Sec.~\ref{Section:OpticalPS}, we proposed to consider the square canonical layout \cite{Clements16} for several reasons, above all because it supports the implementation of any unitary transformation $U$, with no modifications at the hardware level. This is possible because any $m \times m$ $U$ can be decomposed as $U = D \prod\nolimits_b U_b $, i.e., the product of a diagonal matrix $D$ (whose elements are complex and have modulo 1) and $m(m-1)/2$ unitaries $U_b$ implementing complex rotations in the plane spanned by two adjacent modes:
\begin{equation}
 U_b =
\begin{pmatrix} 
    1       & \hdots    & 0         & 0         & \hdots    & 0\\
    \vdots  & \ddots    & \vdots          & \vdots    &           & \vdots\\
    0       & \hdots    & u_{11}    & u_{12}   & \hdots    & 0\\
    0       & \hdots    & u_{21}    & u_{22}   & \hdots    & 0\\
    \vdots  &           & \vdots    & \vdots    & \ddots    & \vdots\\
    0       & \hdots    & 0         & 0         & \hdots    & 1
    \end{pmatrix} \quad,
\end{equation}
\noindent where the $u_{ij}$ ($i,j \in \{1,2\}$) describe the transformation acting on the two modes. This transformation is implemented by means of a pair of phase shifter ($\theta_1$) and imbalanced beam splitter, whose transformation $U^{\textrm{BS}}$ is implemented as a Mach-Zehnder interferometer with tunable phase shifter ($\theta_2$):
\begin{equation}
 U^{\textrm{BS}}(\theta_2) = \frac{1}{2}
    \begin{pmatrix}   
   1 &   i \\    i & 1 
   \end{pmatrix}
     \begin{pmatrix}
   e^{i \theta_2} &   0 \\    0 &   1
   \end{pmatrix}
     \begin{pmatrix}
   1 &   i \\    i & 1 
   \end{pmatrix}.
   \label{eq:MZ}
\end{equation}

Overall, the proposed ansatz can be implemented using well-established, integrated photonic components that only involve tunable phase shifters and fixed, balanced beam splitters (directional couplers).\\
\vspace{3em}


\bibliography{qPSrefList}{}

\begin{thebibliography}{10}
\bibitem{PSisnotNN}
Despite the apparent similarity of their representation, the operation of
  projective simulation (PS), based on the random walk of a single excitation,
  is in stark contrast with that of neural networks: (i) in neural networks one
  has multiple excitations of various magnitude adding up to produce an output;
  (ii) the clip network has arbitrary and dynamic connectivity and (iii) clips
  carry semantics, making PS a genuinely interpretable model.
\end{thebibliography}
\bibliographystyle{ieeetr}


\subsection*{Data availability}

{\footnotesize

Data for all numerical analyses is available at\\ https://github.com/MariusKrumm/PhotonicsPS.

}


\subsection*{Code availability}

{\footnotesize

Code for simulating the quantum episodic and compositional memory and for testing photonic circuits is available at\\ https://github.com/MariusKrumm/PhotonicsPS.

}


\subsection*{Acknowledgements}

{\footnotesize

This project has received funding from the European Union's Horizon 2020 research and innovation program under the Marie Skłodowska-Curie grant agreement, Grants No. 801110 and No. 885567. It reflects only the author's view, the EU Agency is not responsible for any use that may be made of the information it contains. ESQ has received funding from the Austrian Federal Ministry of Education, Science and Research (BMBWF). This work was supported in part by the Austrian Science Fund (FWF) through the SFB BeyondC F7102 and the Volkswagen Foundation (Az:97721).
This work was also co-funded by the European Research Council (ERC) under Project No. 101055129. Views and opinions expressed are however those of the authors only and do not necessarily reflect those of the European Union or the European Research Council. Neither the European Union nor the granting authority can be held responsible for them.

}

\subsection*{Author contributions}

{\footnotesize

F.F., M.K. and H.J.B. conceived the project,
L.J.F., T. M. and H.J.B. contributed to the development of the work.
F.F., M.K. and L.J.F. conceived the analyses related to photonic circuits and to episodic memories, then F.F. carried out the analyses writing a Mathematica software.
M.K. conceived the experiment on the transfer-learning scenario and the variational approach to mimic projective simulation based on loss functions, then carried out the analyses writing a Python software.
All authors contributed to discussing the results and writing the manuscript.

}

\subsection*{Competing interests}

{\footnotesize

The authors declare no competing interests.

}

\newpage


\section*{\textemdash ~ Appendix ~ \textemdash}

\appendix



\section{Package overview}
\label{Section:CompTools}

The above considerations can be tested and explored with a \texttt{Mathematica} package we make available online \cite{Mathematica, Package}. Below we describe its main functionalities.\\

\begin{figure}[t]
	\includegraphics[width= 0.97 \linewidth]{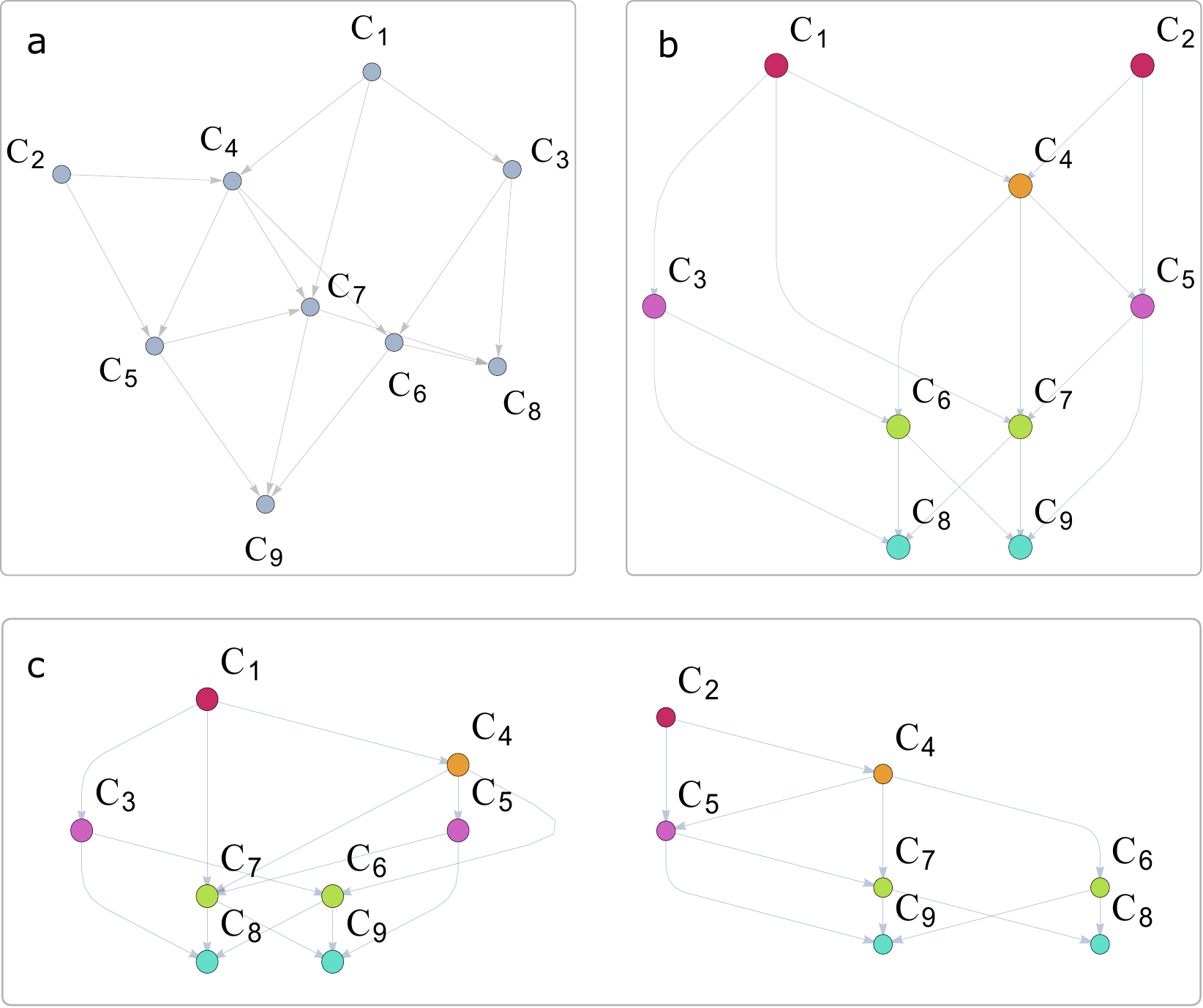}
	\caption{\textbf{Simulating a quantum ECM.}
	a) Directed acyclic graph describing the ECM shown in Fig.~2 in the main text, where each vertex corresponds to a clip $c_k$ (here including percepts and actions).
	b) Vertices, labeled as in panel (a), are rearranged in a layered structure and recolored depending on the order in which the corresponding unitary acts on the excitation (see App.~\ref{Section:SM_C2Qecm} for a discussion on the ordering of the unitaries). Some unitaries that appear consecutively in the ordering can be seen to commute trivially and are thus grouped in the same layer with the same color.
	c) Subgraphs corresponding to percepts $c_1$ and $c_2$. Vertices are colored as in panel (b).      	}
	\label{fig:qECM}
\end{figure}

$\bullet$ \quad Function 1 allows to study the structure of the quantum ECM proposed in this work, starting from the directed acyclic graph (DAG) that describes the connectivity of the ECM. It can be used to test the learning model and to develop extensions. This function operates at an abstract level and, specifically, does not presume an optical implementation. Functions 2-4 below can interact with its output by manipulating the transformations associated with each vertex of the DAG.
Simple examples and details can be found in Fig.~\ref{fig:qECM}.

\begin{algorithm}[H]
   \begin{algorithmic}[1]
    \caption{--- InitializeQuantumECM}
    \Require{\footnotesize The directed acyclic graph (DAG) $G$ describing an ECM.} 
    \Ensure{{\footnotesize An equivalent $G$ with explicit layers; the vertices of $G$ that correspond to percepts and actions; all subgraphs of $G$ that are reachable from each percept (where quantum walks take place); a description of the clip-to-clip connectivity in $G$; extra information.}}
    \label{function:getQECM}
    \end{algorithmic}
\end{algorithm}

$\bullet$ \quad Functions \ref{function:Udec}-\ref{function:getGraph} have been adapted from a package from Ref. \cite{Flamini21}. They make it possible to manipulate unitary transformations on a linear-optical circuit, with a focus on the canonical decompositions. These routines serve to instantiate the abstract representation of a quantum ECM, and allow to study arbitrary circuits as graphs.
They handle the $\vec{\theta}$ parametrization in two formats, one for canonical architectures and one for arbitrary connectivities.

\begin{algorithm}[H]
   \begin{algorithmic}[1]
    \caption{--- UnitaryDecomposition}
    \Require{\footnotesize A unitary $U$ and the choice of an algorithm for its decomposition.} 
    \Ensure{\footnotesize For canonical architectures \cite{Reck94, Clements16}, it decomposes $U$ into phases and transmissivities. If beam splitters are implemented as Mach-Zehnder interferometers, it returns the appropriate phase settings.}
    \label{function:Udec}
    \end{algorithmic}
\end{algorithm}
\vspace{-0.4cm}
\begin{algorithm}[H]
   \begin{algorithmic}[1]
    \caption{--- GetUfromParameters}
    \Require{\footnotesize Parameters in the format handled by Function \ref{function:Udec}, or in a layer-wise format that allows arbitrary connectivity. } 
    \Ensure{\footnotesize Unitary generated by any of three methods with the input parameters  (see Function \ref{function:Udec}). Optional noise and/or losses can be set.}
    \label{function:getU}
    \end{algorithmic}
\end{algorithm}
\vspace{-0.4cm}
\begin{algorithm}[H]
   \begin{algorithmic}[1]
    \caption{--- GetGraph}
    \Require{\footnotesize Information on the mode connectivity in the circuit, in the format handled by Functions \ref{function:Udec} and \ref{function:getU}.}
    \Ensure{\footnotesize Graph representation of the input circuit.}
    \label{function:getGraph}
    \end{algorithmic}
\end{algorithm}

\begin{figure}[t]
    \includegraphics[width = \linewidth]{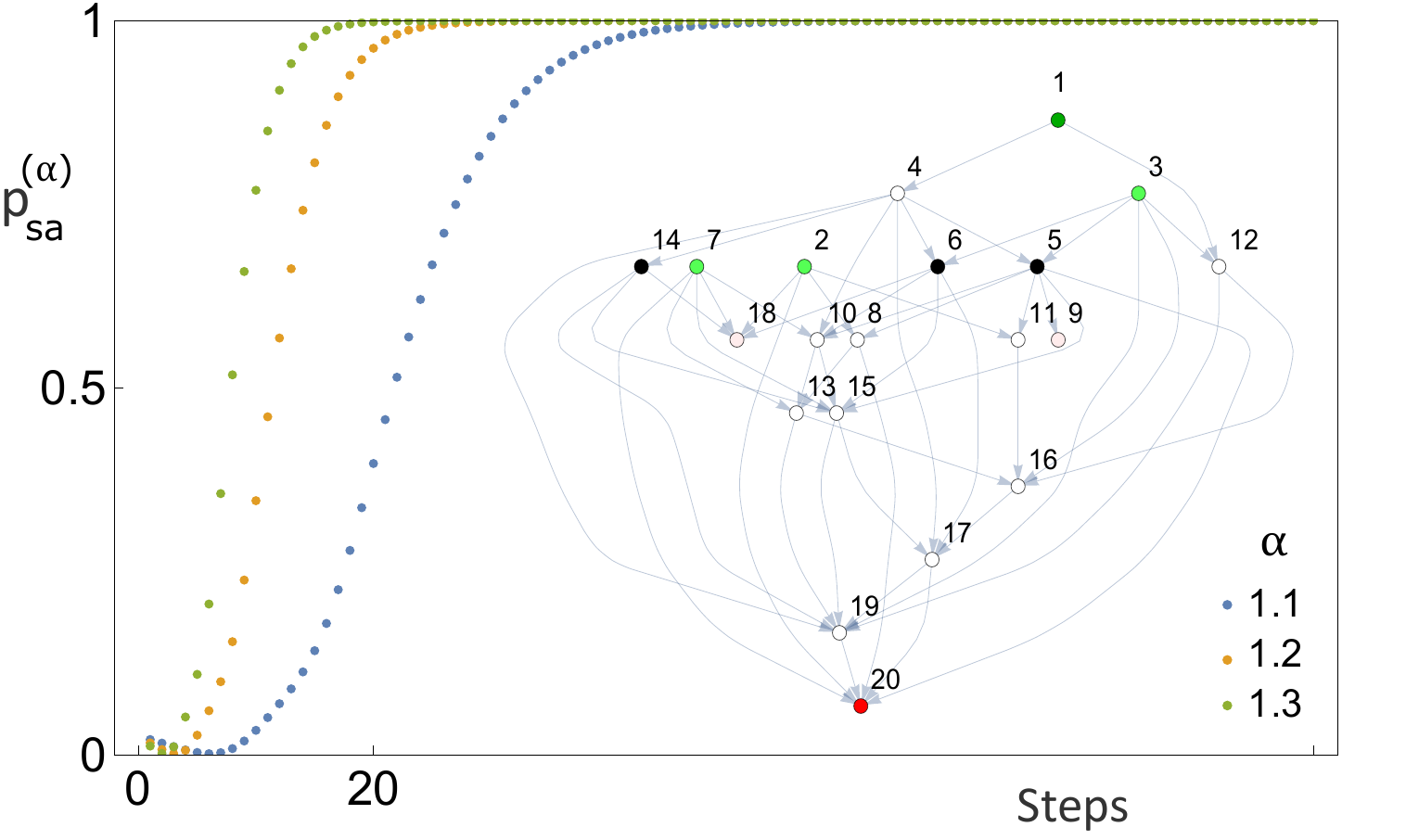}
    \caption{\textbf{Combining causal diamonds and Gram-Schmidt process.}  The Gram-Schmidt orthonormalization can be applied to the $U^k$ associated with an arbitrary path in the ECM. In this example, we reinforce the transition probability $p_{sa}$ by updating the $U^k$ along the longest path (1,4,5,20) that connects a percept ($s$: green, $k=1$) to an action ($a$: red, $k=20$) for the ECM shown in the inset. Here, we assume that at each step the same ($s$, $a$) gets rewarded, for three different factors $\alpha$ (see Sec.~IV in the main text). Inset: The method based on causal diamonds can be extended to the ECM graph. While in the above section the analyses \textit{zoomed in} the structure of each $U^k$, one can also \textit{zoom out} and reinforce $p_{sa}$ by tuning the leaking nodes (``leaking $U^k$'') of ($s$, $a$) along the quantum walk (black vertices, $k \in \{5, 6, 14\}$). Other percepts (actions) are colored in light green (light red).
    }
    \label{fig:CDandGSO}
\end{figure}

$\bullet$ \quad Functions \ref{function:leakingNodes}-\ref{function:CD} enable the update based on the causal diamond for a 2-layer PS (see Sec.~IV B in the main text) and multi-layer PS (see Fig.~\ref{fig:CDandGSO}). They make use of the graph representations given by Functions \ref{function:getQECM} and \ref{function:getGraph}.
Function \ref{function:GSO} implements the update based on the Gram-Schmidt process (see Sec.~IV C in the main text).

\begin{algorithm}[H]
   \begin{algorithmic}[1]
    \caption{--- FindLeakingNodes(AnyDAG) }
    \Require{\footnotesize Graph $G$ describing an optical circuit / a quantum ECM (see Function \ref{function:getGraph}); a percept/action pair ($s$, $a$).} 
    \Ensure{\footnotesize Vertices of $G$ in the causal diamond and leaking nodes of ($s$, $a$).}
    \label{function:leakingNodes}
    \end{algorithmic}
\end{algorithm}
\vspace{-0.4cm}
\begin{algorithm}[H]
   \begin{algorithmic}[1]
    \caption{--- UpdateCausalDiamond}
    \Require{\footnotesize Layered parametrization of the circuit (partially pre-computed for a faster evaluation); a percept/action pair ($s$, $a$); phase shift $\delta \theta$.} 
    \Ensure{\footnotesize New layered parametrization and the corresponding unitary.}
    \label{function:CD}
    \end{algorithmic}
\end{algorithm}
\vspace{-0.4cm}
\begin{algorithm}[H]
   \begin{algorithmic}[1]
    \caption{--- UpdateGramSchmidt}
    \Require{\footnotesize A unitary $U$; a percept/action pair ($s$, $a$); a rescaling factor $\alpha$.} 
    \Ensure{\footnotesize Updated $U$ after one Gram-Schmidt process around $U_{a s}$.}
    \label{function:GSO}
    \end{algorithmic}
\end{algorithm}


\section{Multi-photon quantum walks}
\label{Section:Outlook_manyPercepts}

In this work, we replaced the excitation at the core of the decision making process of PS with a \textit{single-photon} excitation in a quantum walk. In this picture, a photon is always injected into the mode that corresponds to a percept, and the new action is given by the output mode the photon is measured in. An interesting extension of this framework considers the simultaneous injection of \textit{multi-photon} states. A quantum PS agent equipped with this feature could be able to process multiple percepts at the same time, or a percept with a more complex structure. Furthermore, information encoded in $n$ indistinguishable photons injected in $n$ different input modes would interfere at the level of quantum amplitudes, potentially opening up advantages with respect to the classical scheme, where such a feature is currently and independently under investigation. Foreseen potential advantages in this sense are: (i) the advantages inherited by a multi-excitation classical PS; (ii) the exponentially larger (in $n$) percept/action (Hilbert) space; (iii) quantum computational advantages related to the hardness of simulating the dynamics \cite{Aaronson10}.


\section{Quantum walks and frequency encoding}
\label{Section:Outlook_manyFrequencies}

\begin{figure}[t]
    \includegraphics[width = 0.95\linewidth]{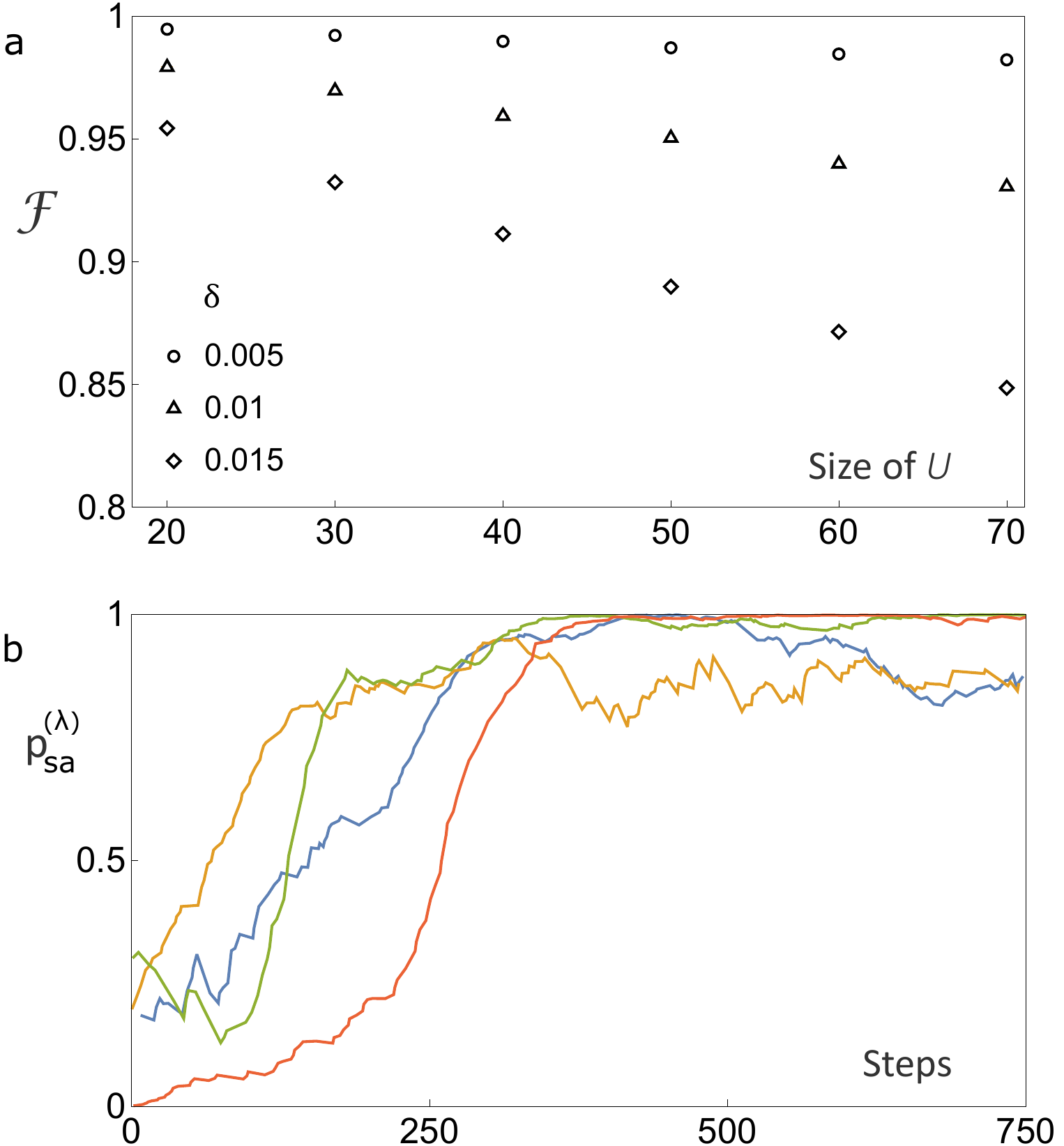}
    \caption{\textbf{Wavelength-dependent quantum walks in the agent's memory.} The wavelength dependence of beam splitters and phase shifters turn a limitation of optical technologies into a potential advantage (see Sec.~\ref{Section:Outlook_manyFrequencies}). a) Fidelity of random \textit{effective} unitary transformations $U$ of increasing size, as perceived by photons with different separation in wavelength $\lambda$ (see legend), averaged over 100 pairs of random unitaries. Each $U$ is numerically generated by uniformly sampling power settings in the phase shifters (see Eq.~\ref{eq:lambdaDependency}) of a square architecture where beam splitters are implemented as tunable Mach-Zehnder interferometers. Error bars (1 standard deviation) are smaller than the point size.
    b) Four learning curves for a 2-layer PS agent based on a $10 \times 10$ square architecture (four rewarded percept/action ($s$,$a$) pairs; each percept/action is encoded in a pair of adjacent input/output optical modes). The figure of merit optimized by the variational approach based on causal diamonds is the geometric mean of the probability $p_{sa}^{(\lambda)}$ for three values of $\lambda$.
    }
    \label{Fig:lambda}
\end{figure}

An interesting possibility offered by an optical implementation is to use additional degrees of freedom of photons, besides the spatial modes, to encode information and/or to train multiple agents in parallel. In particular, frequency appears to be a convenient candidate since the operation of integrated photonic circuits is wavelength-dependent. In fact, both beam splitters' and phase shifters' transformations depend on the wavelength as follows \cite{Flamini15}:

\begin{equation}
 U^{bs}_\lambda \propto
    \begin{pmatrix}   
   \cos{\frac{\pi}{4 \lambda}} &  -\sin{\frac{\pi}{4 \lambda}} \\
   \sin{\frac{\pi}{4 \lambda}} &  \cos{\frac{\pi}{4 \lambda}} 
   \end{pmatrix},
    \quad
     U^{ps}_\lambda(P) \propto
    \begin{pmatrix}   
   1 &  0 \\
   0 &  e^{i \frac{P}{\lambda}}
   \end{pmatrix}
   \label{eq:lambdaDependency}
\end{equation}

\noindent where $P$ is the (suitably rescaled) power setting that controls the phase shifter. With these simplified expressions, we can already get an idea about the average impact of frequency encoding on unitary transformations $U$ of increasing size. Results for this analysis are shown in Fig.~\ref{Fig:lambda}a. We see that, as expected since the architecture consists of hundreds of optical elements, a tiny variation in wavelength $\lambda$ leads to significantly different effective $U$s. This means that photons injected with these deviations would undergo different $U$s, possibly implementing quantum walks in different quantum ECMs.

Finally, we investigate whether a variational approach can work also while additional information is encoded in the frequency degree of freedom. To do so, we simulate quantum learning agents that attempt to solve the same RL task (same rewarded percept/action pairs) using different $\lambda$s. Incidentally, this analysis is similar to, but more sophisticated than, the one presented in Fig.~4d in the main text. Results are shown in Fig.~\ref{Fig:lambda}b, using the approach based on the causal diamond in a 2-layer PS. We see that the agent learns to map each of five percepts to the corresponding rewarded action, at the same time for three different $\lambda$s. Importantly, all phase settings are the same for all $\lambda$, that is, the phase configuration found by the algorithm operates well (as shown) when input with different $\lambda$.

The above considerations suggest that there is a large potential in the use of additional degrees of freedom that influence the quantum walk. Foreseen potential advantages of this aspect are: (i) extra room for more complex RL mechanism; (ii) parallelism in the agent's decision making process.


\section{Numerical details of the transfer-learning scenario}
\label{Section:SM_DetailsTransfer}

In this section, we report details on the numerical analysis described in Sec.~V in the main text, performed with PyTorch.

\begin{figure}[h]
    \includegraphics[width = 0.85\linewidth]{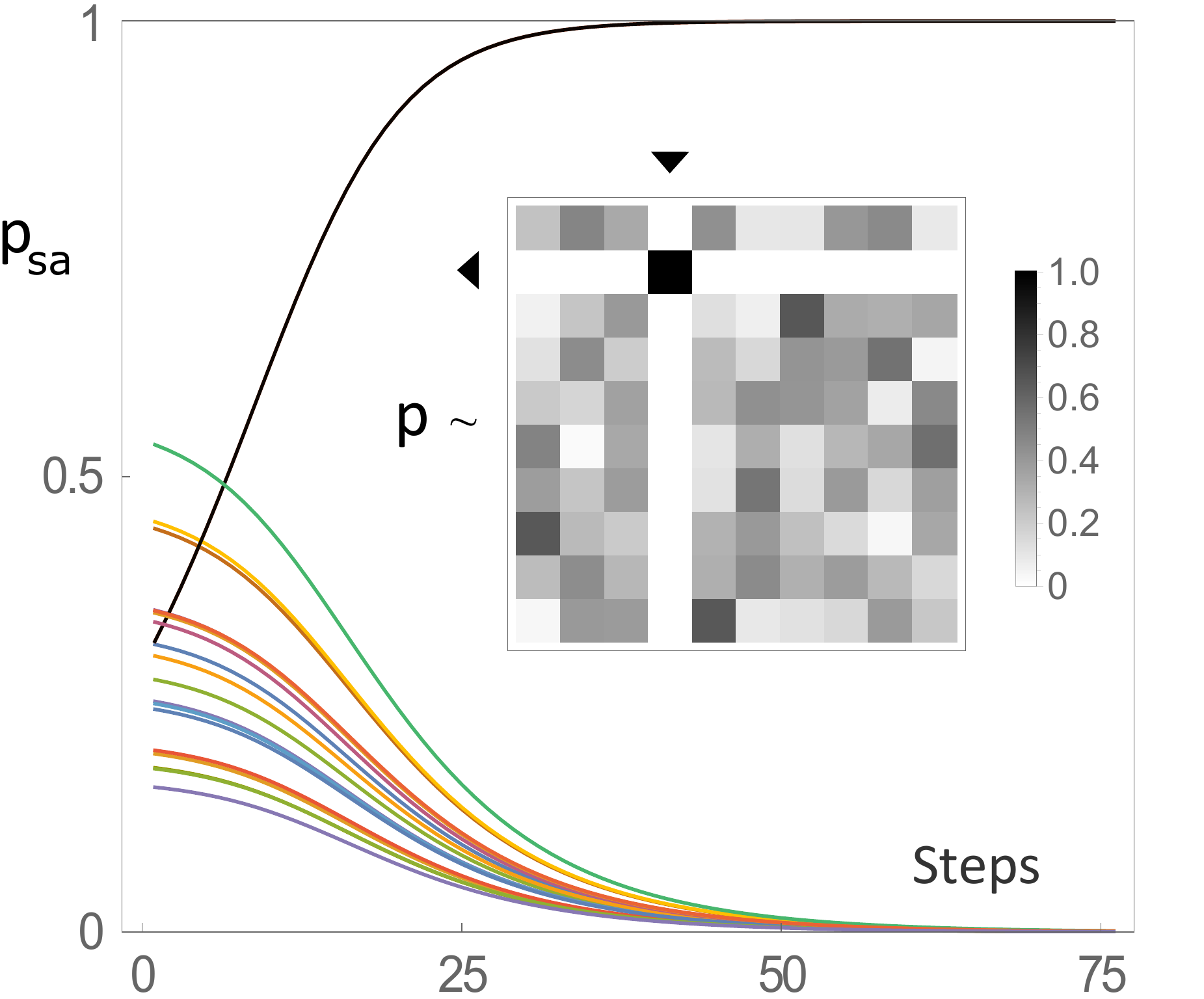}
    \caption{\textbf{A transfer-learning scenario for PS agents.} 
    a) Structure of the ideal, classical ECM considered in the transfer-learning scenario. Panel adapted from Ref.~\cite{Ried19}.
    b) Percepts $s$ consist of the values of three observables which characterize particles.
    The experiment consists of two stages. (i) In the first stage, the PS agent is set up to learn a middle layer of nodes, in which each node corresponds to one observable and one value of that observable. (ii) In the second stage, this middle layer is kept fixed. The task layer is exchangeable and it represents one yes/no question about pairs of observables. Since the excitation of the classical ECM is localized, in the middle layer the classical agent can only use the value of one observable. Interference allows the quantum PS agent to combine knowledge of several observables at once. 
    c) Layout of the quantum PS agent. To fulfill the unitarity constraints, the connection from percept to middle layer is an interferometer tree with $9$ outputs (one tree per percept). The connection from the middle layer to the task layer is a $9\times 9$ square architecture, shared by all percepts but different for each task layer.
    Two output modes correspond to the yes/no answers, the others are discarded.
    d) Prediction accuracy $A_Q$ of the quantum PS agent for each of the 27 experiments (yes/no question). Inset: Learning curve for a single experiment (dark orange). A classical agent can achieve at most an accuracy $A_C=8/9$ (horizontal line, for both plots), always lower than $A_Q$.}
    \label{fig:transferLearning}
\end{figure}

In Ref. Eva et al.~\cite{Ried19}, the authors considered three observables, each of which can take three values. We label these observables and their values $0,1,2$. Each percept can be written as $(v_0, v_1, v_2)$ with $v_j \in \{0,1,2\}$, and $v_j$ is the value of observable $\mathcal O_j$. In total, there are $3^3 = 27$ percepts. In addition to the percept layer, there is a middle layer and a final layer (see Fig.~\ref{fig:transferLearning}). The purpose of the middle layer is to explicitly represent the value of each observable. More specifically, this layer contains $3^2 = 9$ nodes, each corresponding to one observable-value pair. For each percept $(v_0, v_1, v_2)$, the idea is that the agent learns to visit the middle clips $(\mathcal O_j, v_j)$ with uniform probability $\frac{1}{3}$, while not visiting the other middle clips $(\mathcal{O}_j, v)$ corresponding to wrong values $v$ of the observables.

For the proposed quantum PS, we represent the correct middle-layer state of percept $(v_0 , v_1 , v_2)$ as $\frac{1}{\sqrt 3} \left( \ket{v_0} + \ket{3 + v_1}+\ket{6+v_2} \right) $. For example, $(2,1,2)$ is represented by $\frac{1}{\sqrt 3} \left( \ket{2} + \ket{4} + \ket{8} \right) $.
Since these states can be non-orthogonal for different percepts, for each percept we use a separate binary tree (see Fig.~\ref{fig:transferLearning}b) with 9 output leaves (27 such trees in total), instead of the unitaries $U^k$ shown in Fig.~2 of the main text.

To make a fair comparison to a classical PS agent, we consider the following training. Each of the $27$ percepts is sampled with uniform probability. Then the middle layer is measured to output one (observable, value) pair. Depending on whether the pair is predicted right or wrong, a fixed reward $\pm r$ is given. We use the simplified loss function of Eq.~(11) without forgetting-mechanism ($\gamma  = 0$), and only reward the current percept-action pair ($\eta  =1$).
Since we use separate binary trees for each percept, and the goal of the tree is just to prepare one simple state, the middle layer can be trained to achieve the intended representation essentially perfectly. 

After the middle layer is trained, we fix its weights, meaning that we will not adjust the binary trees for anything that follows. Now, we introduce the final layer, which we call the task layer. The task layer represents yes-no-questions about the values of observables. Our transfer learning scenario requires that the task layer be exchangeable, while we keep the middle layer fixed and independent of the task layer. The task layers we consider here pick two observables $\mathcal O, \mathcal O'$ and ask whether these observables have specific values $v_{\mathcal O}$ and $v_{\mathcal O'}$. There are $27 = \binom{3}{2} \cdot 3^2$ such task layers.
The quantum PS agent uses a 9x9 square architecture for each task layer, as shown in Fig.~\ref{fig:transferLearning}c. We emphasize that the same task unitary is used for each percept. Only two output modes are used, which correspond to yes/no-answers.
The $27$ percepts are sampled uniformly.
Fig.~\ref{fig:transferLearning}d shows the achieved (weighted) accuracies for all two-observable task layers.

Both classical and quantum PS agents have three layers: \emph{percept}, \emph{middle} and \emph{task} layer. While in classical PS each percept $s$ is represented by a vertex in the ECM graph, in this analysis the quantum PS agent associates with it a (linear-optical) binary tree \cite{Flamini20}, to overcome the limitation due to the unitarity constraints in the middle layer. All its phases are initialized to $\frac{\pi}{4}$. The classical PS agent is expected to pick an observable uniformly at random, and then send the excitation to the middle clip with the right value. For the quantum PS agent, we wish that the percept $(v_0, v_1, v_2)$ is represented by the state $\frac{1}{\sqrt{3}} \big( \ket{v_0} + \ket{3 + v_1} + \ket{6 + v_2} \big) $. To this end, we first train the binary trees, considering a RL scenario where percepts are sampled i.i.d. uniformly. We use the simplified loss function $\mathcal{L}_{\mathrm{PS}}$ of Eq.~(11) in the main text, with the Kullback-Leibler divergence applied to the probability distribution $\{p_{sa} , 1-p_{sa}\}$. Furthermore, we use $\mathrm{ReLU}$ functions to cut off the target values therein such that they are in $[0,1]$:
\begin{align}
    p^{(t+1)}_{sa} = 1.0 - \mathrm{ReLU}\Big(1.0 - \mathrm{ReLU} (p^{(t)}_{sa} + r)\Big)
\end{align}
As a reward, we assign $r := \pm 0.1$ depending on whether or not the agent predicts the right value.
To ensure that each observable is picked uniformly, we implement a curiosity mechanism by including the Shannon neg-entropy of the observables in the loss function: 
\begin{align}
    \mathcal{L}_{\mathrm{Shannon}} = \log(3) + \sum_{j = 0}^2 p_{s O_j} \log p_{s O_j}
\end{align}
where $p_{s O_j}$ is the probability of the $j$-th observable $O_j$, given a percept $s$.
To enforce that the relative phases among the modes of the middle layer vanish, we also add to the loss function the $\ell_1 $ norm of the complex phases, i.e. $\mathcal{L}_{\mathrm{phase}} : =\sum_{m=0}^8 |\Phi_{s m}| $. Here, $\Phi_{s m}$ is the phase in the middle-layer mode $m$ when percept $s$ is considered.
The full loss function for the middle layer is then:
\begin{align}
    \mathcal{L}_{\mathrm{full}} = 1.0 \ \mathcal{L}_{\mathrm{PS}} + 10.0 \ \mathcal{L}_{\mathrm{Shannon}} + 1.0 \ \mathcal{L}_{\mathrm{phase}}
\end{align}
We use experience replay, and add up the loss functions of a batch of $600$ sampled percepts before running 10 optimizer steps (Adam optimizer \cite{Kingma14}) and wiping the batch. The learning rate is set to $0.01$, decreased to $0.001$ once the accuracy of the last batch was at least $0.95$. 
Training stops when the accuracy of the middle layer exceeds $0.99$ for $10$ consecutive rounds and, at the same time, the $\ell_1$-norm of the phases weighted by $p_{sm}$ is smaller than $0.1$. Practically, the optical elements are simultaneously updated by computing the gradients of the sum of loss functions for many ($s$, $a$, $r$)-tuples, once the replay sample memory has reached a threshold. This means that there is no immediate update after a reward $r$ but only after several tuples.

Once the middle layer is trained, we fix its phase shifters so that the trees do not change anymore. Now, we introduce the task layer, which is exchangeable and independent of the middle layer. While Eva et al. only considered single-observable experiments, our task layers pick two observables ($\mathcal O_0, \mathcal O_1$) and ask whether they have specific values $v_{\mathcal O_0}$ and $v_{\mathcal O_1}$.

The quantum PS agent uses one 9x9 interferometer for each task layer.
Again, we use experience replay, with a batch size of $500$ (much larger than the number of percepts) and train the agent for $40000$ rounds. Since our tasks ask for specific values of two observables simultaneously, only $\frac{1}{9}$ of the percepts will correspond to a yes-answer. Hence, in addition to the raw accuracy, we use a weighted accuracy where yes-answers are given an additional weight $8$ (otherwise the agent could just always answer no, which would yield a seemingly good accuracy of $\frac{8}{9}$). The loss function is the one in Eq.~(8) in the main text, with a reward equal to $\pm 0.1$ for a right or wrong answer, and an additional factor $8$ if the right answer is \emph{yes}, the Kullback-Leibler divergence for $\{ p_{sa} , 1-p_{sa} \}$, $\mathrm{ReLU}$ functions to cut off the target values and neither glow nor forgetting. We normalize the probability of the \emph{yes}- and \emph{no}-mode of the task layer to one, i.e. we post-select on getting one of these two modes. For the learning rate and number of optimizer steps we use the same settings as before.

Overall, the quantum PS agent consistently achieves over $0.97$ accuracy in all two-observable tasks (as shown in Fig.~6 in the main text).


\section{Classical-quantum equivalence of ECM graphs}
\label{Section:SM_C2Qecm}

In this section we formalize the connection between the directed graph of a classical ECM and the quantum walks that derive from it. We refer to this set of quantum walks (from input to output) as \emph{unitary routing}. We begin by introducing some basic elements from graph theory.

\begin{definition}[\bf{Topological ordering}]
    A topological ordering $o_G$ of a directed graph $G = (V,E)$ is a map, $o: V\rightarrow \{1,\dotsc, |V|\}$, which defines a total ordering of vertices of $G$ such that for every directed edge $(v_j, v_k)$ from vertex $v_j$ to vertex $v_k$, $v_j$ comes before $v_k$ in the ordering, i.e., $o(v_j)< o(v_k)$.	
    Let $\mathcal{O}_G$ be the set of all topological orderings of $G$.
\end{definition}

\begin{figure}[b]
    \centering
    \includegraphics[width = 0.6 \linewidth]{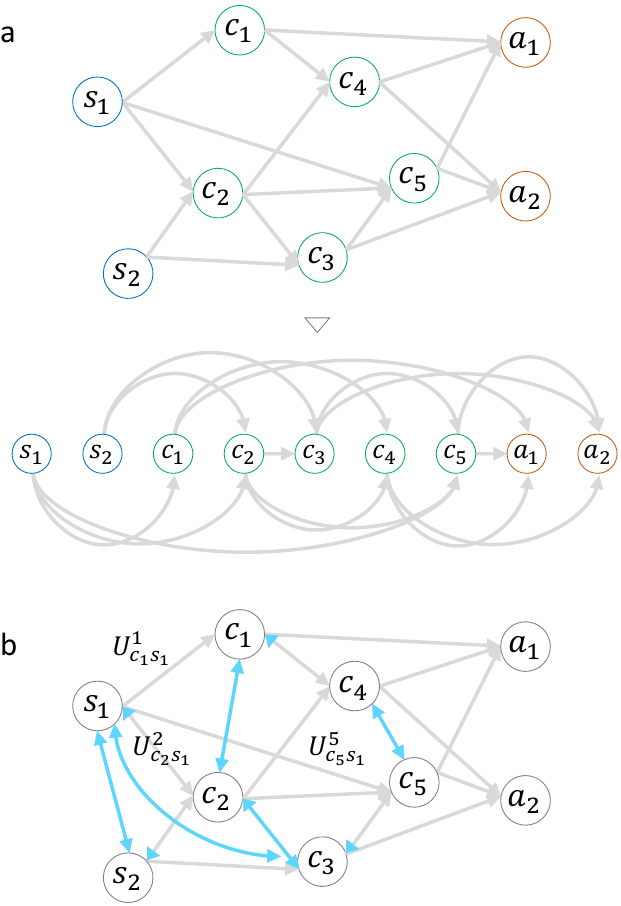}
    \caption{a) Directed acyclic graph shown in Fig.~2 in the main text (top) and one of its topological orderings (bottom). Notice that, for the sake of simplicity, all vertices had already been labeled with subscripts in ascending order. b) Unitarity constraints in the mode-mixing transformations $U^k$ in Eq.~\eqref{fromGtoRouting} may generate edges (blue) that are not present in the classical ECM.}
    \label{fig:app:graphECM}
\end{figure}

\noindent An example of topological ordering is presented in Fig.~\ref{fig:app:graphECM}.

Directed acyclic graphs (DAGs) are directed graphs that do not contain loops. They can be defined as follows \cite{Bang-Jensen09}:

\begin{definition}[\bf{Directed acyclic graph}]
	A directed graph is acyclic if and only if it admits a topological ordering.
\end{definition}	

\noindent Thus, the graph in Fig.~\ref{fig:app:graphECM} is an example of a DAG. It is worth noticing that the topological ordering of a DAG such as in Fig.~\ref{fig:app:graphECM} is not unique, because swapping consecutive vertices (in the topological ordering) which are not connected by an edge yields another valid topological ordering.

Below we recall two additional, useful definitions:
\begin{definition}[\bf{Ordered DAG}]
	An ordered DAG $\vec{G}$ is a DAG $(V,E)$ with $V = \{1, \dotsc, |V|\}$ such that $o:V\rightarrow \{1, \dotsc, |V|\}, o(j) = j$ is a topological ordering of $(V,E)$.
	
	For any DAG $G = (V,E)$ with topological ordering $o$ we denote the induced ordered DAG by
	\begin{align}
		\vec{G}^{(o)} =&  \Big(\{o(v_i) \ | \ v_i\in V\}_i, \\
       &  \{(o(v_j),o(v_k)) \ | \ (v_j,v_k)\in E\} \Big). \nonumber
	\end{align}
Let $\mathrm{ocDAG}$ be the set of all ordered and connected DAGs.
\end{definition}

\begin{definition}[\bf{Parents and children}]
	Let $G = (V,E)$ be a graph. For any vertex $v_i\in V$, let $P_G(v_i)$ be the set of parents of $v_i$, i.e., $P_G(v_i) = \{v_j|v_j\in V, \exists (v_j,v_i)\in E\}$. For any vertex $v_i\in V_G$, let $C_G(v_i)$ be the set of children of $v_i$, i.e., $C_G(v_i) = \{v_j|v_j\in V, \exists (v_i,v_j)\in E\}$.
\end{definition}

Equipped with the above definitions, we proceed by defining an ECM graph. Here we focus on ECMs described by connected DAGs.
\begin{definition}[ECM graph]
    An ECM graph is a connected DAG $G = (V,E)$ whose vertices are called either percepts, $\mathcal{S}  = \{v_j|v_j\in V, P_G(v_j) = \emptyset\}$, actions $\mathcal{A}  = \{v_j|v_j\in V, C_G(v_j) = \emptyset\}$, or intermediate clips $\mathcal{C}  = V\backslash (\mathcal{S}\cup\mathcal{A})$.
\end{definition}

The quantum version of a classical ECM graph is defined via a unitary routing. We refer to Table \ref{table:routes} for an overview of the terminology introduced in this section.

\begin{table}[t]
    \begin{center} 
        \begin{tabular}{p{0.2\linewidth}p{0.75\linewidth}} 
        Object        & Intuition    \\ 
        \midrule
        Routing         & Sequence of parametrized unitary transformations, constructed from a graph $G$.  \\
        Route          & A specific sequence of the routing, where all parameters have been fixed.    \\
        Router         & The map from $G^o$ to its routing   \\
        \bottomrule
    \end{tabular}
    \caption{Summary of the main quantities introduced in App.~\ref{Section:SM_C2Qecm}.}
    \label{table:routes}
    \end{center}
\end{table}

\begin{definition}[\bf{Unitary route}] \label{eq:def:route}
    Let $G = (V,E)$ be an ECM graph and $o$ a topological ordering of $G$. A unitary route of the induced ordered DAG $\vec{G}^{(o)}$ is a sequence of $|V|\times |V|$ unitaries, ordered in ascending order of their superscripts,
    \begin{align}
    \mathrm{route}(\vec{G}^{(o)}) = \left(U^{o(v_i)}\right)_{v_i\in \mathcal{C}\cup\mathcal{A}},
    \end{align}
    \noindent whose matrix elements are given by
    \begin{align}
	U^i_{jk} = \begin{cases}
		u^i_{jk} &\textrm{ if }j, k\in \Big( \{i\}\cup P_{\vec{G}^{(o)}}(i) \Big) \\
		\delta_{jk} &\textrm{ else,} \label{fromGtoRouting}
	\end{cases} 
    \end{align}
    where the $u^i_{jk}\in\mathbb{C}$ are only subject to unitarity constraints.
\end{definition}    

For example, with reference to the example in Fig.~\ref{fig:app:graphECM} and the topological ordering $(o(s_1),...,o(a_2))=(1,...,9)$, $P_{G} (c_2)=(s_1, s_2)$ and $P_{\vec{G}^{(o)}}(4)=(1,2)$, hence 
\begin{equation}
    U^4 =
    \begin{pmatrix} 
        u^4_{11}    & u^4_{12}           & 0       & u^4_{14}  & 0   & \cdots  & 0 \\
        u^4_{21}           & u^4_{22}    & 0       & u^4_{24}  &     &         &   \\
        0           & 0           & 1       & 0         &     &         &   \\
        u^4_{41}    & u^4_{42}    & 0       & u^4_{44}         & 0   &         &   \\
        0           &             &         & 0         & 1   &         & \vdots  \\
        \vdots      &             &         &           &     & \ddots  & 0   \\
        0           &             &         & 0         & \cdots    & 0       & 1 
    \end{pmatrix}  .
\end{equation}

\begin{definition}[\bf{Unitary routing}] \label{eq:def:Routing}
    Let $\mathrm{routing}(\vec{G}^{(o)})$ be the set of all unitary routes of $\vec{G}^{(o)}$, generated by the freedom in choosing the coefficients $u^i_{jk}$ within the constraints of unitarity. Let also $\mathrm{ocROUTING} = \{\mathrm{routing}(\vec{G}^{(o)})\}_{\vec{G}^{(o)}\in\mathrm{ocDAG}}$.
\end{definition}

\begin{definition}[\bf{Unitary router}] \label{eq:def:router}
    We define a \emph{unitary router} as the function
    \begin{align}
    \mathrm{router}: \;     & \mathrm{ocDAG} \rightarrow \mathrm{ocROUTING},\\ \nonumber
                            & \vec{G}^{(o)}\mapsto \mathrm{routing}(\vec{G}^{(o)}).
    \end{align}
\end{definition}

Given a unitary route $\mathrm{route}(\vec{G}^{(o)})$, the unitary matrix describing the whole quantum ECM is then given by 
\begin{align}
    U = \prod_{U^i\in \mathrm{route}(\vec{G}^{(o)}) } U^i,
\end{align}
where the product is ordered from left to right in descending order of the superscripts. Note that here the superscript $i$ refers to the position of the vertices in the ordering $o$, while in the main text the superscript refers to an unspecified ordering, and that in general the $U^i$ do not commute.

If an ECM graph $G$ admits more than one topological ordering, $G$ can have different unitary routings. Nevertheless, there is the following one-to-one quantum-classical correspondence:

\begin{lemma}[\bf{Quantum-classical correspondence}]
The function $\mathrm{router}$ is bijective.
\end{lemma}
\emph{Proof}. To show bijection we will first show injectivity, that is, we must show that $\mathrm{router}(\vec{G}^{(o)}) = \mathrm{router}(\vec{H}^{(o')})\Rightarrow \vec{G}^{(o)} = \vec{H}^{(o')}$. According to the definition of routing, a $\vec{G}^{(o)}$ has $\mathrm{routing}(\vec{G}^{(o)})$  only if (i) it has vertices $V = \{1, \dotsc, n\}$, where $n$ is the number of rows (or columns) of each $U^i$ in any of the routes in $\mathrm{routing}(\vec{G}^{(o)})$, and (ii) it has directed edges $E = \{(j, k) \ | \ \exists\, \mathrm{route}(\vec{G}^{(o)})\in\mathrm{routing}(\vec{G}^{(o)})\textrm{ and }\exists\, U^k\in\mathrm{route}(\vec{G}^{(o)}) \textrm{ such that } U^k_{jk}\neq 0 \}$. This completely determines $\vec{G}^{(o)}$ as $\vec{G}^{(o)} = (V,E)$. Thus $\mathrm{router}(\vec{G}^{(o)}) = \mathrm{router}(\vec{H}^{(o')})$ can only be true if $\vec{G}^{(o)} = \vec{H}^{(o')}$.

Surjectivity is trivial because the codomain of $\mathrm{router}$ is defined as its range, which completes the proof that $\mathrm{router}$ is bijective. $\square$

\vspace{1em}

While a one-to-one correspondence exists on the level of ordered DAGs $\vec{G}^{(o)}$ and unitary $\mathrm{routings}$, there is an interesting freedom on the level of ECM graphs and/or unitary routes.
In the latter case, a unitary route is not sufficient to uniquely identify a $\vec{G}^{(o)}$ because if $u^i_{jk} = u^i_{kj} = 0$ for some $j\neq k$, the corresponding directed edge $(j,k)$ in $\vec{G}^{(o)}$ cannot be recovered just from the unitary route. Instead, one would recover a $\vec{G}^{(o)'}$ without that edge. Classically, this corresponds to the case of a vanishing $h$-value, $h_{jk} = 0$.
In the former case, it is worth noticing that ECM graphs are defined without ordering. However, ordering matters for unitary routes since in general the $U^i$s do not commute. This represents an interesting departure from classical ECMs: different orderings can give rise to different unitary routings.


\section{Notes on leaking nodes}
\label{Section:SM_LeakingNodes}

In this section we discuss in more detail the update mechanism based on causal diamonds. The applicability of this method goes beyond the scope of this work, and can find use with any linear-optical interferometer.\\

\begin{figure}[h]
    \centering
    \includegraphics[width = 0.75 \linewidth]{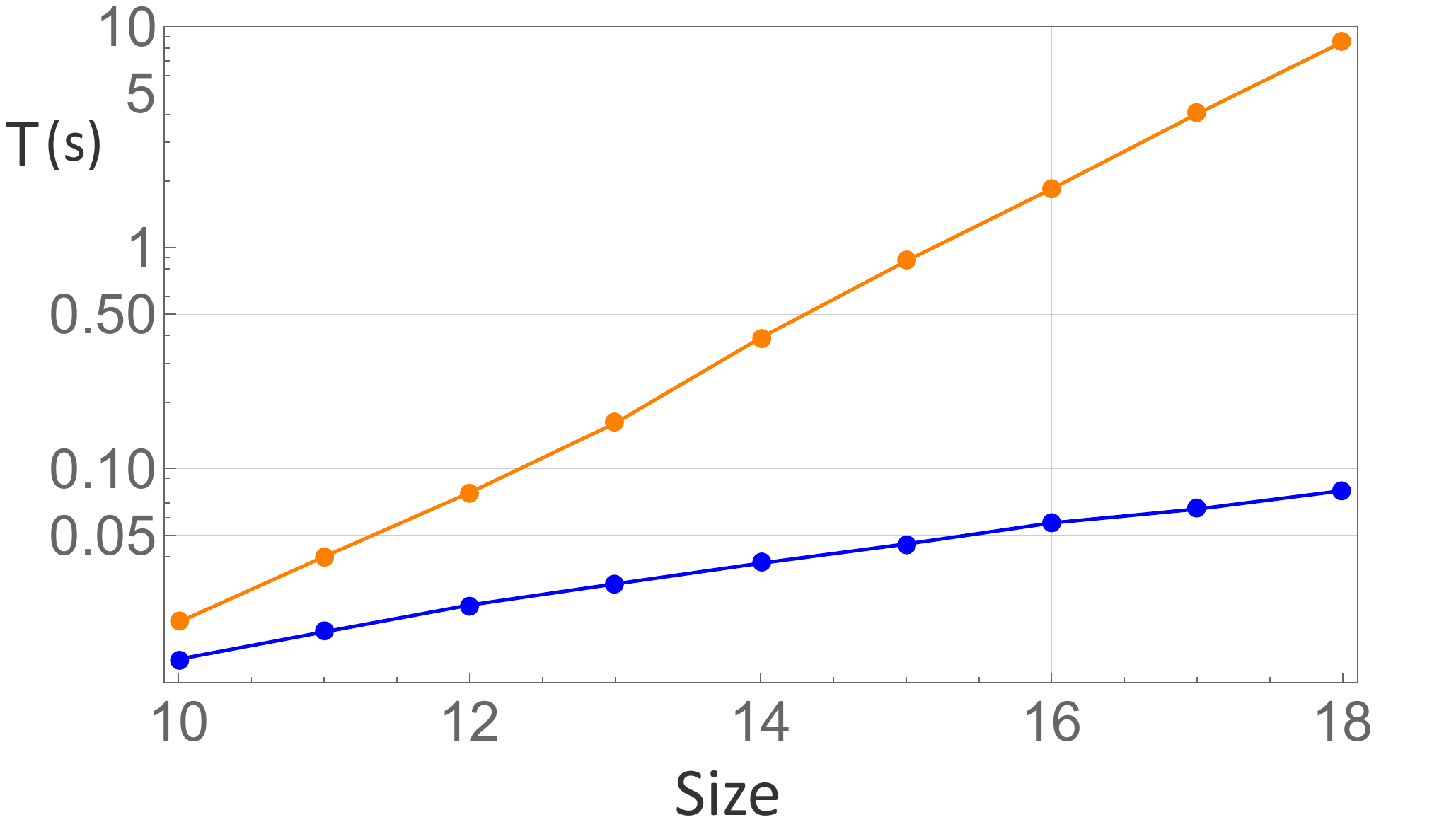}
    \caption{Wall-clock time $T$ required by a naive algorithm (orange) and the proposed algorithm (blue) to find the leaking nodes in circuits of different sizes, on a standard laptop. The naive algorithm scans all paths from the input mode to the output mode (exponentially many) and checks for signatures that a path is at the boundary of the causal diamond. The refined algorithm scans all vertices of the causal diamond, checking whether or not all vertices that can be reached in one step still belong to the causal diamond. Notice that only the scaling and the approximate value of $T$ are relevant in this plot.}
    \label{fig:app:1}
\end{figure}

\textit{Identifying leaking nodes \textemdash}
Leaking nodes are vertices of the directed graph that describes the circuit connectivity, with the property that light can leak out of the causal diamond between input and output mode (hence never reaching the output). There are several ways to identify such nodes. The most intuitive way is by enumerating all paths from input to output, keeping track of spatial information (e.g., coordinates on a grid) that allows to identify the boundaries of the causal diamond. However, this approach may not be convenient for more complex or three-dimensional circuits, and the number of paths scales exponentially with the number of optical modes. Here, we tackle this problem by observing that leaking nodes have a simple property: One outgoing edge connects to a node that is not in the same causal diamond. Since the list of nodes in the causal diamond is supposed to be already known (easy to retrieve by graph search), this approach scales much more favorably than the above alternatives (see Fig. \ref{fig:app:1}), and requires no computations over paths.\\

\textit{Updating leaking nodes \textemdash}
Updating circuits using the causal diamond is conceptually simple: (i) one starts from the layered description of an $m \times m$  optical circuit, (ii) tweaks one by one each phase shifter, according to a predefined mechanism, and (iii) multiplies the unitary transformations $U_l$ associated with each layer $l$ to rebuild the full transformation.
The issue with this approach is that it entails an unnecessary computational overhead, in that all but one phase shifters are kept fixed during the update. Hence, one can store the expression of the partial unitary transformation until ($U_{\rightarrow l}$) and after ($U_{l \rightarrow}$) the current layer (not included), and update them as we move on to the next phase shifter:

\begin{align}
    \textup{Layer update: \;} \begin{cases}
	U'_{\rightarrow l} = U_l \; U_{\rightarrow l} \\
	U'_{l \rightarrow} = U_{l \rightarrow} \; U^{-1}_l
    \end{cases}. \label{eq:updateCDLayerwise}
\end{align}

\begin{figure}[ht!]
    \centering
    \includegraphics[width = 0.99 \linewidth]{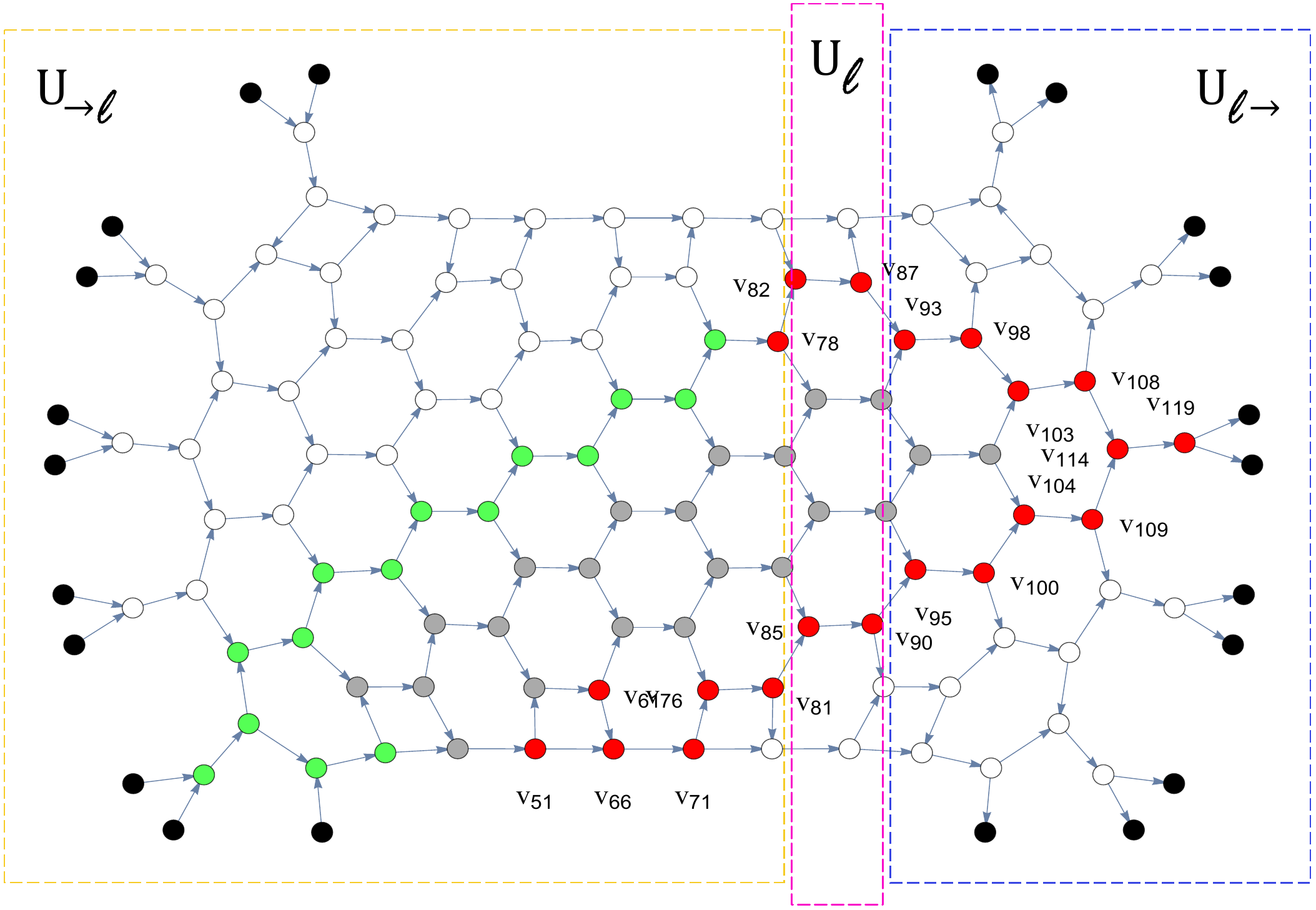}
    \caption{Sketch of the subroutine introduced to update circuits using the causal diamond (here: green, gray, and red nodes), when this is not performed via gradient descent on the loss function. Once the subset of nodes to be updated (red) is identified, we tune them layer by layer (here, fuchsia frame). All layers of optical elements before and after the current one are treated as two separate unitary transformations ($U_{\rightarrow l}$ and $U_{l \rightarrow}$, respectively). Whenever a layer is updated, moving to the next layer only requires the update of the two partial transformations, according to Eq. \ref{eq:updateCDLayerwise}.}
    \label{fig:app:2}
\end{figure}

\noindent Notice that $l$ runs over the indices of layers in the causal diamond, so that, for example, $U_{\rightarrow l}=I$ only when the input and output modes are along the diagonal of the circuit. An illustrative example of this strategy is shown in Fig. \ref{fig:app:2}, where the fuchsia box tracks the current layer and, starting from the leftmost red node, it separates the two partial transformations $U_{\rightarrow l}$ (yellow) and $U_{l \rightarrow}$ (blue).
Within the shared code and on a standard laptop, this approach is tens to hundreds of times faster than a naive simulation of the entire circuit, depending on the size of the circuit.\\

\begin{figure*}[h]
    \centering
    \includegraphics[width = \linewidth]{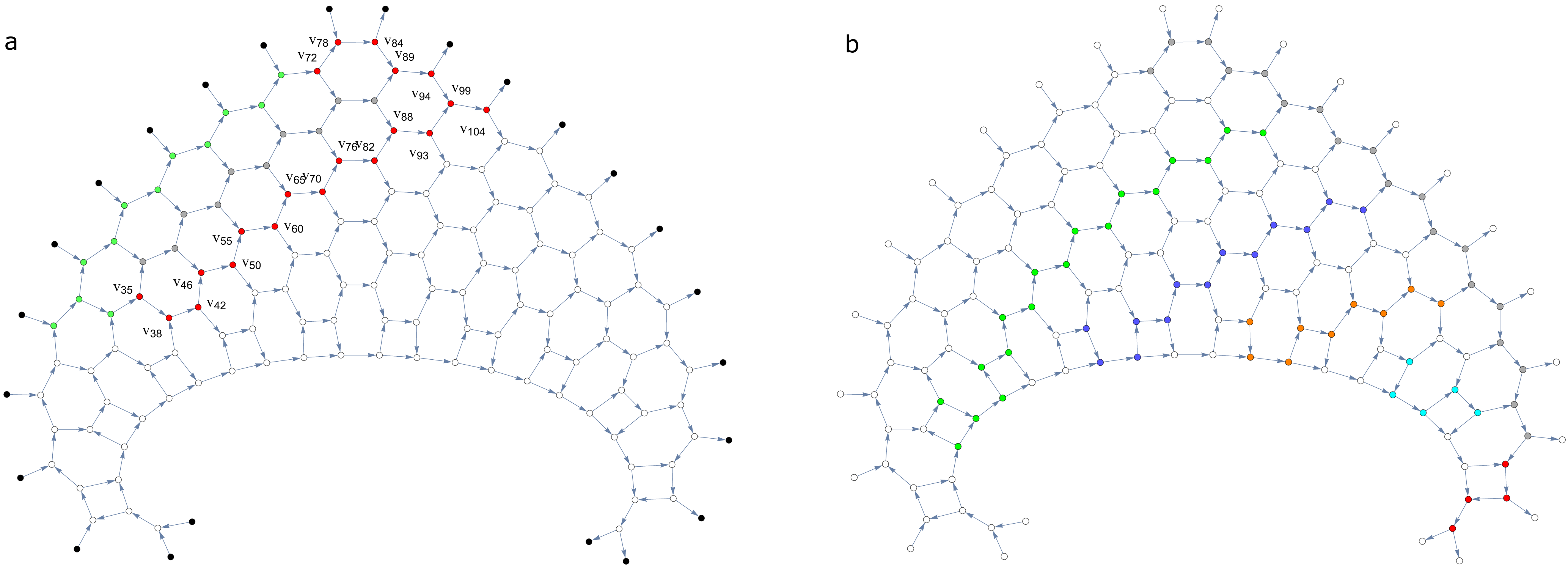}
    \caption{Here we extend the analyses carried out on the square layout \cite{Clements16} to the triangular layout \cite{Reck94}, for a single (a) and multiple (b) rewarded input/output pairs.
    While this work focuses on the square layout for its symmetry, balanced losses and greater modularity, readers using the triangular layout can find this technique of interest also for applications beyond the scope of this work.}
    \label{fig:app:3}
\end{figure*}

\begin{figure*}[h]
    \centering
    \includegraphics[width = 0.95 \linewidth]{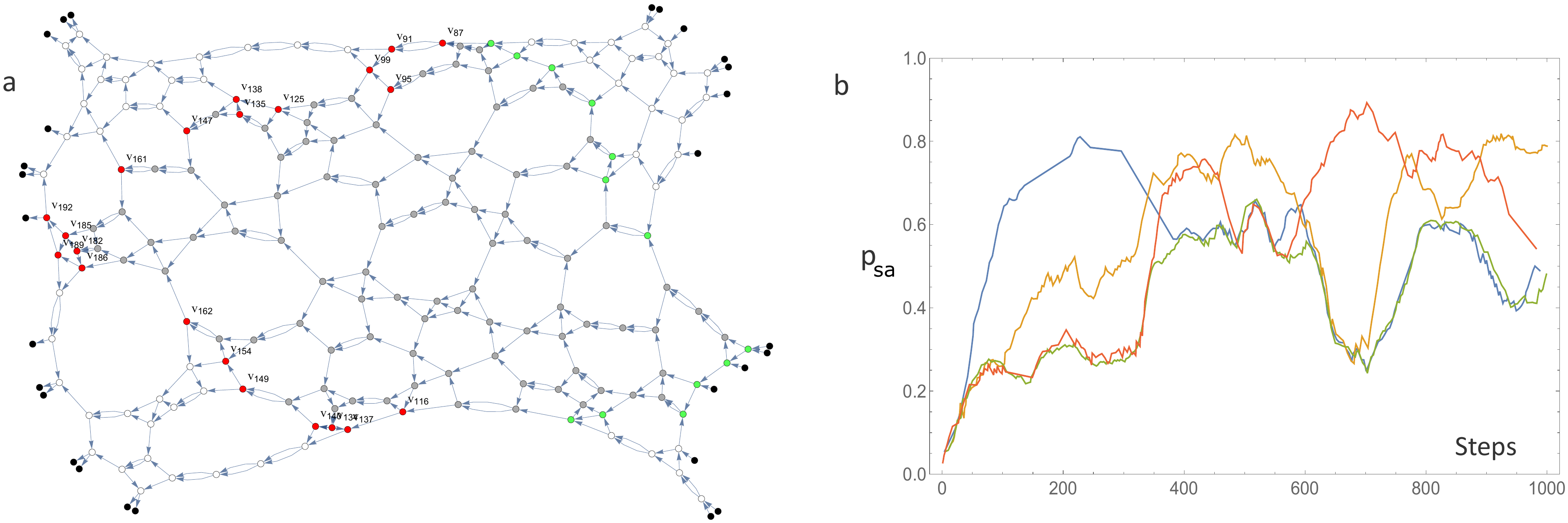}
    \caption{a) The analyses related to the causal diamond can be extended to optical circuits with arbitrary mode connectivity. b) Learning curves for a toy problem using the random optical circuit shown in panel (a). In this example there are four rewarded, random input/output pairs, corresponding the the four curves in panel (b). Compared to the case of square and triangular architectures, here the learning curves are degraded due to the lower number of independent nodes between in/out paths and the emergence of bottleneck nodes, which are hard to optimize for multiple paths.}
    \label{fig:app:4}
\end{figure*}

\textit{Causal diamonds in other layouts \textemdash}
The proposed framework to quantize and implement PS on photonic circuits uses the square architecture to support single-photon quantum walks \cite{Clements16}. This notwithstanding, many of the ideas related to the causal diamond still apply to other layouts. For instance, causal diamonds can be identified also in the triangular architecture \cite{Reck94} (see Fig. \ref{fig:app:3}), and the layer-wise update mechanism can be applied also here. Similarly, one can use the proposed ideas and code to study causal diamonds with arbitrary, even random mode connectivity (see Fig. \ref{fig:app:4}a), and still achieve decent learning curves for multiple input/output pairs (see Fig. \ref{fig:app:4}b). While this result may not have an immediate application for PS, the opportunity may be relevant for more complex optical circuits that go beyond the scope of this work.\\

\end{document}